# A REVIEW OF COMPLEX SYSTEMS APPROACHES TO CANCER NETWORKS

Uthamacumaran, A.

a_utham@live.concordia.ca

Concordia University, Montreal

**ABSTRACT**

Cancers remain the lead cause of disease-related, pediatric death in North America. The emerging field of complex systems has redefined cancer networks as a computational system with intractable algorithmic complexity. Herein, a tumor and its heterogeneous phenotypes are discussed as dynamical systems having multiple, strange attractors. Machine learning, network science and algorithmic information dynamics are discussed as current tools for cancer network reconstruction. Deep Learning architectures and computational fluid models are proposed for better forecasting gene expression patterns in cancer ecosystems. Cancer cell decision-making is investigated within the framework of complex systems and complexity theory.

**INTRODUCTION TO COMPLEXITY**

Recent advances in our understanding of cancer – using modern tools of complexity theory – has revealed that cancer is a complex adaptive system. Many emergent properties of cancer, such as heterogeneous clonal expansion, replicative immortality, patterns of longevity, rewired metabolic pathways, altered ROS (reactive oxygen species) homeostasis, evasion of death signals, hijacked immune system, self-sufficient growth signals, and metastatic invasion are all indications of cancer's *complex adaptive* nature (Hanahan and Weinberg, 2011 [1]).

A complex system is a nonlinear dynamical system of many interacting parts which adaptively respond to the perturbations of their environment (Shalizi, 2006 [3]; Ladyman and Wiesner, 2020 [2]). The signatures of a complex adaptive system include nonlinearity, emergence, self- organized patterns, interconnected multi-level structures, critical phase-transitions, computational irreducibility, unpredictability and multi-scaled, feedback loops. In simple terms, the concerted whole cannot be defined by the sum of its interacting parts (Wolfram, 1988 [4]; 2002 [5]; Gros, 2011 [6]). Complex systems are also chaotic – moving or modifying small sections or pieces can have unexpected consequences.

The irreducible systemic makeup of cancers makes them impossible to understand without complex systems theory. Complex systems theory, also known as complexity science, emerged from dynamical systems theory in the 1960s. It is an evolving interdisciplinary study of how the interactions between the various parts of a system give rise to its collective behaviors. Complex systems resurged with the amalgamation of computational complexity theory, a sub-branch of computer science which studies the solvability of problems pertaining to systems. Computational complexity theory, or simply denoted here as complexity theory, studies the resources such as time, space, and algorithms needed to solve computational problems. Complexity theory and complex systems theory are often interchangeably used within different disciplines of science. The division between the two will be blurred here as the study of complex systems such as cancer networks heavily depend on the algorithms derived from complexity theory. A clear example to illustrate the union of the two are cellular automata. Elementary cellular automata consist of a one-dimensional lattice of cells with Boolean states, where each cell is updated by looking at its neighboring cells and itself. Cellular automata demonstrate that simple computer programs following discrete rules can produce highly complex behaviours (Wolfram, 1984 [7]). Such discrete dynamical systems have been shown applicable in the study of complex systems such as self-replicating structures, nonequilibrium pattern formation, fluid dynamics and cancer growth (Wolfram, 2002 [5]; Jiao and Torquato, 2011 [8]; Monteagudo and Santos, 2012 [9]).

Complexity theory attempts to provide a computational description of Nature. Intractability, computational undecidability and irreducibility, the bread and butter of complexity theory are not easily understood with mathematical equations (Johnson and Garey, 1979 [10]). Early in the 1930s, mathematical logicians found that determining whether a mathematical statement was true or false had a fundamental limitation (Sipser, 1997 [11]). While Turing and Church developed the theory of computing, Gödel's incompleteness theorems showed that no recursively-axiomatizable formal system can fully encode true mathematical statements.  The limits of algorithmic solvability were questioned. The question of which problems are efficiently computable was first informally addressed in a letter by Gödel to von Neumann (1956). The P vs. NP problem asks, are problems to which solutions are quickly verifiable, easily solvable? In complexity theory, problems are categorized by a complexity class, the time it takes an algorithm to efficiently solve them as a function of size. While P denotes the class of computational problems that are solvable in polynomial-time, NP (Nondeterministic Polynomial-time) problems are quickly checkable but are either intractable or solved by brute-force searching. That is, the exact solution of an NP problem grows exponentially (or factorially) with the size of the number of elements in the system. Exact solutions are generally unattainable and require some form of approximating rules (heuristics). To illustrate, consider the following analogy. Is searching necessary to find a needle in a haystack? This depends on our tools. If a magnet was available, the vast space of possibilities need not be explored to find the needle.

Throughout this review, some specific examples of cancer-related problems that necessitate the use of algorithms apparatus from complexity theory are discussed. The reconstruction of cancer networks from gene expression data will be discussed as a computational complexity problem. Finding attractors/repellors in the state-space of these networks is an NP-hard problem where attractors may characterize the distinct cellular phenotypes in a tumor sample. The article intends to shift the current thinking of cancers as stable equilibria or fixed-point attractors towards that of strange attractors. While fixed-point attractors correspond to cancer phenotypes acting as points fluctuating in a valley or apex of the epigenetic landscape, strange attractors are points which aperiodically fluctuate back and forth between certain hills/valleys but remain bound to this abnormal pattern in state space. A valley corresponds to a lower entropy indicating a differentiated cell state, whereas an apex (hill) on the landscape corresponds to a high entropy (chaotic fluctuations) indicating a higher stem cell potential. If validated, finding the minimal set of genes that converts a strange attractor to a fixed-point attractor may present a solution to reprogramming cancer stem cells to benignity. The various subsections illustrate that cancer dynamics consist of chaotic, nonequilibrium behaviors at multiple scales. Algorithms that may be able to capture strange attractor behaviors in cancer networks are proposed.

**STEM CELL REPROGRAMMING**

Reprogramming terminally differentiated cell fates to stemness was an intractable problem until the pioneering works of Nobel Laureates Shinya Yamanaka and John Gurdon. Today, the Yamanaka factors can be replaced by a cocktail of small molecules with higher reprogramming efficiency (Hou et al., 2013 [12]). The chromatin remodelling by the chemical factors allows the facilitated binding of Yamanaka transcription factors (TFs) and overcome the epigenetic barriers to dedifferentiate cells to iPSC (induced Pluripotent Stem Cell) states. Since then, many new algorithms have been discovered for chemically altering the epigenetic landscape of differentiated cell states and thereby minimizing their trajectories towards stem cell attractors (Rais et al., 2013 [13]; Ranquist et al., 2017 [14]; Hernandez et al. 2018 [15]). For example, micro-RNA-based reprogramming generated iPSC clones claimed with up to 90% efficiency in human fibroblasts (Kogut et al., 2018 [16]). There was > 200- fold increase in iPSC reprogramming efficiency when the culture media contained antagonists of TGF-$\beta$, MEK/ERK (mitogen activated kinases) inhibitors and thiazovinin (Rock inhibitor) (Saito et al., 2019 [17]).

Reprogramming cancer stem cells to healthy iPSC states remains an intractable problem and debated since the molecular heterogeneity of cancer stem cells (CSCs) varies in time by patient, tissue-type and microenvironmental cues (O'Brien-Ball and Biddle, 2017 [18]). Nevertheless, limited findings indicate CSCs can be reprogrammed in vitro to iPSC-like states or chemically directed towards particular cell-fate commitments. For example, human pancreatic cancer cells were iPSC reprogrammed in vitro using episomal vectors and demonstrated a lack of tumorigenicity (Khoshchehreh et al., 2019 [19]). Another example, neuroblastoma cells were reprogrammed directly into osteoblastic lineage by the mTOR inhibitor, Rapamycin (Carpentieri et al., 2016 [20]). However, in time, the reprogrammed cells can still revert to malignancy or generate novel cancer phenotypes.

While stemness denotes the unlimited self-renewal and differentiation properties of immortalized cancer cells, the definition of a cancer stem cell is ambiguous. Cancer stem cells (CSCs) were first isolated in AML (Acute Myeloid Leukemia) where surface markers distinguished a subpopulation of cells that had tumor-initiating capacity (Bonnet and Dick, 1997 [21]). In highly fluid cancers such as leukemia, distinct cancer stem cells are identified. For example, the 17-gene stemness score (LSC17 score) is a prognostic biomarker used for assessing acute myeloid leukemia (AML) relapse in clinical care (Shlush et al., 2014 [22]; Ng et al., 2016 [23]). However, the question remains openly debated whether all cancer cells of a tumor are potentially stem cells (Battle and Clevers, 2017 [24]). Dynamic behaviors such as the interconvertibility of cell fates (trans-differentiation), de-differentiation and quiescence add further layers of complexity to the cancer stem cell identity. Current findings suggest the phenotypic plasticity of cancer stem cells (CSCs) are dynamic, microenvironment- dependent and less constrained than believed. Identifying the master Gene Regulatory Networks (GRN) coordinating cancer stemness remains burdensome and thus, a roadblock for reprogramming cancer stem cells to benignity.

Epigenetic dysregulation is a hallmark of cancer which allows their adaptation to fluctuating environments. The epigenetic burden is greater in pediatric cancers than in adults (Capper et al., 2018 [25]; Filbin and Monje, 2019 [26]). Moreover, there are multiple complex programs such as the transfer of exosomes and microRNAs involved in the cancer stemness problem. For instance, by regulating the PI3K/AKT/mTOR signaling pathway, miR-126 confers leukemic stem cells' self-renewal, quiescence and therapy resistance (Lechman et al., 2016 [27]). In addition, recent findings demonstrate cancer cells form highly complex, bidirectional feedback loops with healthy cells. For example, high-grade gliomas integrate into the electrical networks of healthy neurons wherein depolarizing potassium currents promote glioma progression (Venkatesh et al., 2019 [28]). Despite the complexity, several patterns in cancer stem cell networks have been recognized in the past decade indicating possible routes towards their cell fate reprogramming to benignity. To illustrate, a few examples of critical regulatory pathways and epigenetic patterns identified in highly morbid, pediatric brain cancers are given.

Sox2, a Yamanaka factor, is an essential driver of cancer stem cell sub-populations in Glioblastoma Multiforme (GBM) (Suvà et al., 2013 [29]; 2014 [30]). PI3K/mTOR (mammalian Target of Rapamycin) and MEK/ERK pathways were shown to be critical to the self-renewal of glioma stem cells (GSC) and mediate cancer stemness in brain tumors (Sunayama et al. 2010 [31]). Regardless of the divergent clonal evolution of such tumors, key driver mutations in histone H3 post-translational modifications and IDH1/2 (isocitrate dehydrogenase) were tractable patterns of epigenetic reprogramming in GBM stem cells (Suvà et al., 2014 [30]; Salloum et al., 2017 [32]). The recurrent somatic driver mutations in H3 histones are identified as distinct molecular patterns in pediatric GBM, leading to amino-acid substitutions at key histone residues such as K27M, K36M, and G34V/R. Progression to higher grade glioma displayed an overall decrease in methylation, and hypermethylation of a small subset of CpG islands associated with developmental regulators, including FOX, SOX and TBX family genes. The epigenetic reprogramming may arrest cells into a permanently self-renewing state (Bai et al., 2016 [33]). TERT promoter/telomerase mutations often occurred later for rapid growth and relapsed tumors, indicating a critical attractor required for cancer immortalization (Korber et al., 2019 [34]; Stead and Verhaak, 2019 [35]).

Single-cell transcriptomics-based cell lineage reconstruction revealed four distinct targetable clusters in pediatric medulloblastoma (Northcott et al., 2017 [36]; Vladoiu et al., 2019 [37]). Apart from the differential gene expression signatures, CpG island methylation profiles were also distinct between the clusters as shown by t-SNE (t-distributed stochastic neighbor embedding) plots. Transcriptomics revealed glial-tumor interactions regulate brain metastases via upregulated EMT/MET (Epithelial-Mesenchymal Transition) pathways. The EMT pathways coordinate chemical pattern formation (i.e., morphogenesis) and confer dynamic switching in cancer stem cell fates (Wingrove et al., 2019 [38]).

The epigenetic landscape of GBMs (Glioblastoma Multiforme) shows tremendous spatiotemporal heterogeneity. However, a core set of neurodevelopmental transcription factors (POU3F2, SOX2, SALL2, OLIG2) were identified to be essential for GBM propagation and stemness (Suvà et al.,2014 [30]). As seen, single cell multi-omics are indicating a link between the Yamanaka factors and the epigenetic reprogramming of cancer stem cells. CRISPR screens identified stemness and chemotherapy resistance regulators in patient-derived glioblastoma stem cells including members of the SOX gene as well (Macleod et al., 2019 [39]).

Increasing evidence suggest that cancer stemness is a plastic state acquirable by all cancer cells (Gong et al., 2019 [40]). Depending on the tumor microenvironment (stem cell niche), 'terminally differentiated' cancer cells can dedifferentiate and acquire stem cell properties or transdifferentiate into other cancer phenotypes. It has been speculated that solid tumor cells exposed to chemotherapeutic agents can dedifferentiate to a plastic stem cell fate and interconvert to other cancer cell types. The speculations were recently confirmed, where cancer cells exposed to chemotherapy were shown to promote tumor recurrence and aggressiveness as result of the increased cancer stemness (Xiong et al., 2019 [41]).

Temozolomide (TMZ), an oral chemotherapy drug, is often the first-line treatment for Glioblastoma. The findings showed that temozolomide not only increased the glioma stem cell population, but also reprogrammed CD133-negative glioma cell lines and patient-derived, differentiated glioma cells into GSC stem cells in part by the activation of HIF (Hypoxia-Inducible Factors) and iPSC networks (Yamanaka factors). GBM stemness and recurrence were also seen in exposure of differentiated glioma cells to ionizing radiation therapy. Collectively, these reviewed findings suggest that despite the tremendous success rates in most pediatric cancer therapies with current regiments (ignoring the life-long side effects and quality of living), current treatments can induce the cellular reprogramming of differentiated cancer cells into plastic stem cell fates, and thereby causes therapy resistance, disease progression and cancer recurrence (Xiong et al., 2019 [41]).

**BOOLEAN NETWORKS**

In this section, an overview of the mathematical models currently employed in reconstructing cancer networks is presented. Waddington (1942 [42]) first described cellular development as an energy landscape where cell fate bifurcations are visualized as balls rolling up hills and down valleys (attractors) (Bhattacharya et al., 2011 [43]; Wang et al., 2011 [44]). The attractors may correspond to the distinct cell fates. The transitions between the attractors are considered as random walks on a network (Perkins et al., 2014 [45]). Gene expression is regulated in a combinatorial fashion by the diffusion-kinetics of chemical structures and proteins such as transcription factors. Due to the stochastic nature of molecules, the changes in molecular concentrations of transcription factors are generally defined by partial and stochastic differential equations (Davila-Vederrain et al., 2015 [46]; Elowitz et al., 2002 [47]). To simplify the system, it is often assumed the synthesis-degradation dynamics of the mRNA and its protein products settle to steady-state probability distributions (Chen et al., 1999 [48]; Swain, 2016 [49]). The dynamic interactions between these molecular regulators and chemical species in a cell that governs the gene expression levels of mRNA and their protein products is known as a Gene Regulatory Network (GRN). Gene regulatory networks and protein-protein interaction (PPI) networks usually exhibit scale-free topologies where the degree of regulatory connections follows the power law distribution (Perkins et al., 2014 [45]). Scale-free networks have a distinct characteristic in their structure, where the most-connected networks form hubs while the remainder are sparsely connected (Barabási and Oltvai, 2004 [50]; Zhu et al., 2007 [51]). Many other network topologies such as those generated by the Watts-Strogatz model (small-world networks) and Barabási-Albert model (preferential attachment) algorithms may also be observed in metabolic and signaling networks (Zhu et al., 2007 [51]).

Given a Gene Regulatory Network, there are two general approaches to their reconstruction. The classical approach is studying its nonequilibrium dynamics via the Master equation (Cross and Greenside, 2009 [52]). The GRN dynamics is represented by a Chemical Master Equation (CME) that describes the time evolution of the probability distribution of discrete molecular quantities. However, the CME is in practical cases non-integrable and often requires mean-field approximations where steady-state protein distributions are assumed (Cao and Grima, 2018 [53]). Complex networks can exhibit non-trivial topologies and dynamical structures.

The Fokker-Planck equation is the derivation of the Master equation used to characterize the time-evolution of chemical diffusion systems underlying the regulatory networks, while Hill equations, Michaelis-Menten kinetics, nonlinear transport equations and differential rate laws account for the gene-protein interactions (Metzler et al., 1999 [54]). Chaotic dynamics can emerge in the higher-dimensional mapping of such networks beyond a critical threshold in gene expression fluctuation (Mestl et al., 1996 [55]). However, solving these differential equations become intractable once the number of chemical species in the system reaches more than three (Swain, 2016 [49]). That is, an interacting system of as little as three variables can exhibit deterministic chaos. The time-evolution of a cell state's probability density in state-space is given by the Fokker-Planck equations:

$$\frac{\partial \rho}{\partial t} = -\sum_i \frac{\partial}{\partial x_i}[A_i(x)\rho] + \frac{1}{2}\sum_{i,j} D_{ij}(x)\frac{\partial^2}{\partial x_i x_j}\rho$$

where, D is the diffusion matrix, A is the drift vector characterizing the drag and random forces encountered in Brownian motion, x is the spatial coordinate in one dimension and $\rho$ is the probability density. In a general 1-D setting, the mean path is determined from a path integral assuming a Gaussian noise term as given by: $\rho(x,t) = \int \boldsymbol{Dx}\, exp[-S(x)] = \int \boldsymbol{Dx} \exp\{-\int L(x(t))dt\}$, where the sum over all possible paths contributes to the cell fate trajectory. The diffusion coefficient D is often given by the Stokes-Einstein relation (assuming a spherical molecule's diffusivity in a quiescent fluid), S is the action and L is the Lagrangian or weight of each path such that the probability flux $\frac{\partial \rho}{\partial t} + \nabla \cdot J = 0$ is conserved (Li and Wang, 2013 [56]; Wang, 2015 [57]). Assuming steady-state solutions, the energy potential at a given point in the landscape is $V(x) = -\ln \rho(x)$ and corresponds to a cell state's energy density. However, $\frac{\partial \rho}{\partial t} \neq 0$ in dissipative and open, nonequilibrium systems, as is the case for biological systems.

In contrast, the Gene Regulatory Network can be represented as a graph. The use of machine learning algorithms provides an alternate route to study cancer networks without the fine details of the differential equations involved in their diffusion-kinetics. However, the union of both approaches alone can provide a more detailed mechanistic representation. Various algorithmic approaches exist for combining the principles of information theory and graph-theoretic networks to study gene expression dynamics. Rather than solving the Master equation or inferring differential rate laws from kinetic experiments, complex systems approaches such as Stochastic Simulation Algorithms, Dynamic Bayesian networks, Monte Carlo methods, Boolean networks and piecewise deterministic Markov processes (PDMP) to name a few, are often used to simulate gene expression dynamics (Palmisano and Priami, 2013 [58]; Herbach et al., 2017 [59]; Lin et al., 2018 [60]).

Graphs are abstract mathematical structures that allow us to model the relationships between elements forming complex dynamical networks. A graph G is $G = (V, E)$, where V is the set of all vertices (nodes) and E is the set of all edges (links connecting the nodes). Each vertex represents an element such as a gene or a protein and the edges are the regulatory interactions between the elements, forming a network. There are several key features describing the network architecture. These include the degree, directionality, clustering coefficient, shortest (optimal) path length, closedness, betweenness and algorithmic complexity (Zhu et al., 2007 [51]). In a directed graph, each vertex has an in-degree (edges coming into the vertex) and out-degree (edges going out from V) creating flow networks (Barabási and Oltvai, 2004 [50]). Network properties are deduced from the network topology, the way in which the nodes and edges are arranged. A flow network is a directed graph where each edge has a capacity and each edge receives a flow. A path along the network is a sequence of edges that begins at a vertex of a graph and travels along edges of the graph, connecting pairs of adjacent vertices. Weights can be assigned to the edges representing the distance, time and travel cost of flows networks (e.g., normalized gene expression values). The minimum cost flow problem and source localization algorithms are most pertinent in the signaling network reconstruction. For instance, the Dijkstra's algorithm is often used to find the single-source shortest path between two nodes. Identifying sub-structures within a network such as the master regulatory modules of a gene-regulatory network can become computationally difficult as the network complexity increases (Barabási and Posfai, 2016 [61]; Barabási and Oltvai, 2004 [50]).

In 1969, Stuart Kauffman [62] proposed a Random Boolean network (RBN) model of Gene Regulatory Networks (GRN); i.e., a discrete dynamical system wherein a gene can either be on or off, represented as binary occupancies of 0 or 1. The RBN model is one of many complex systems approaches to study gene expression dynamics. Rather than solving the Master equation of a GRN, Boolean Networks provide a simplistic rule-based description of gene regulatory networks. The concentration levels in many regulatory processes behave as a Hill-function, giving rise to a sigmodal curve that can be approximated as a dichotomous step-function justifying the Boolean binarization of gene expression. Numerous Boolean models have succeeded the RBN, including continuous analogs and hybrid models of Boolean networks. For instance, kinetic Monte Carlo methods can simulate biological networks in Boolean state space (Stoll et al., 2012 [63]).

The dynamics of Boolean Networks can be represented in directed state graphs. Each regulatory component is represented by a node of the graph. The directed edges between these components represent their regulatory interactions that are expressed by Boolean functions (Schwab et al., 2020 [64]). In most approaches to Boolean networks, binarization of high-throughput gene expression data is required to infer the Boolean functions. The state of a Boolean Network at one point in time *t* is defined by a vector X(t). Considering all possible combinations of N-genes gives a network state-space of $2^N$ possible states. The state transition from a state X(t) to its successor state X(t + 1) can be obtained by: a synchronous update where all Boolean functions are applied at the same time, or as an asynchronous update, where only one randomly chosen function is updated per step. A trajectory through the state graph describes the networks' behavior over time. Attractors represent the long-term behavior of these state graphs and may characterize cellular phenotypes. When using the asynchronous update, complex attractors can emerge (Schwab et al., 2020 [64]).

The dynamics of Boolean Networks are divided into three regimes depending on the structure of their state space: ordered, critical, and chaotic. The *criticality* of GRNs has been discussed as fluctuating at the phase-transition point between ordered and chaotic regimes for the dynamics of those networks. When perturbations are introduced, ordered GRNs are so robust that they just sustain existing cellular functions (Kim and Sayama, 2018 [65]). In the ordered state, any moderate perturbation is rapidly dampened and the network returns to its original attractor. On the contrary, in chaotic GRNs, the attractor cycles are very long and exhibit unpredictability due to sensitive dependence on initial conditions (i.e., slightly different initial states lead to exponentially diverging trajectories in state-space). In between, critical GRNs can simultaneously withstand perturbations and generate new attractors.

Kauffman's NK automaton consists of N genes (Boolean variables) each able to regulate K neighboring genes in a directed graph and associated with a Boolean function. A second-order phase transition is observed in RBNs indicating the transition between order and chaos, as given by the relationship:

$$K_C = \frac{1}{2p_c(1-p_c)}$$

where the subscript c denotes critical value, K is the input connectivity, and Boolean functions are chosen at random with an average of $2^K p$ true entries, where $p \in [0,1]$. The dynamics of an RBN is known to be determined by $K$, namely, $K < K_c$ corresponds to the ordered phase, fixed-point attractors and cyclic/periodic attractors and $K > K_c$ denotes the unstable (chaotic) regime. $K = K_c$ corresponds to the edge of chaos/criticality (Glass and Hill, 1998 [66]). Derrida and Pomeau (1986) [67] demonstrated that to take two random initial configurations in the RBN and measure their overlaps using the normalized Hamming distance is analogous to finding Lyapunov exponents in the continuous dynamics. At K > 2, the Hamming distance between two initially close attractor states grows exponentially in time on average, denoting phase-transition to chaotic dynamics. Alternately, Luque and Solé (2000) [68] used the concept of a Boolean derivative to define the Lyapunov exponents in RBNs as: $\lambda = \log[2p(1-p)K]$, where if $\lambda < 0$ corresponds to an ordered phase, $\lambda > 0$ indicates the chaotic phase and, if $\lambda = 0$ corresponds to criticality.

As previously stated, inferring parameters from gene expression data to fit into the differential equations characterizing the chemical diffusion-kinetics is a challenging problem because these equations do not have analytic solutions. Furthermore, the time-course gene expression data are usually sparse and noisy (Cao et al., 2012 [69]). Distinguishing chaotic gene expression from noise remains a major problem in systems biology. To overcome some of these challenges, hybrid Boolean networks can be formulated as a coarse-grained limit of the differential equations of the system. There are even extensions to the RBN model where the nodes can take more than two-values, forming multi-valued networks. The state and time-continuous dynamics of gene regulation at transcription sites is often considered as a delay-differential equation. The logical structure of Boolean networks can be implemented into these time-delay differential equations. The hybrid models can be tuned for the kinetic rates of transcription factors, protein synthesis-degradation rates and the kinetics of gene bursting from experimental data at different time points. For example, BooleanNet is a Python package available for designing biological networks (Albert et al., 2008 [70]). Synchronous and asynchronous updating strategies are available in this algorithm with an extension to hybrid models by piecewise differential equations.

Consider a simple regulatory gene circuit represented by the reaction logic: $A + B \to C$, the delay differential equation becomes:

$$\frac{d[C(t)]}{dt} = \frac{\alpha[A(t)][B(t)]}{k + [A(t)][B(t)]}(t-\tau) - \gamma[C(t)]$$

The probability of transcription factors A and B, binding to gene promoters to transcribe product C (mRNA) is shown here as a first-order autonomous chemical reaction (i.e., assuming no autocatalysis, self-inhibition or negative-feedback loops for simplicity) (Buchler et al., 2003 [71]; 2005 [72]). Here, $\alpha$ denotes the transcription rate at the promoter at full activation, k is the reaction rate constant and $\gamma$ is the degradation rate of the product C. In Boolean mapping, the above-defined continuous differential equation's state and time-discrete dynamics is given by $C(t), A(t-1)$ and $B(t-1)$ taking binary values of: $\{0,1\}^N$ to represent off/on states.

Gene networks can also be modelled by embedding a logical switching network into Piecewise linear differential equations. When the threshold function in the differential equations are replaced by a step-function, piece-wise linear equations are obtained. The equations were shown to be appropriate models for systems in which there are switch-like interactions between the elements, such as those in Boolean and Hopfield networks (Glass and Pasternack, 1978 [73]). Although in principle, chaotic behaviors can emerge in asynchronously updated Boolean networks, robust algorithms for the detection of strange attractors remains inadequate (Glass and Pasternack, 1978 [73]).

From a complex systems approach, one must ask: are we limiting gene expression dynamics by casting a model-system into a differential equation? For instance, are epigenetic modifications, chromatin remodelling or diffusion-mediated chemical turbulence accounted for in these models? Epigenetic modifications (e.g., DNA methylation, acetylation rates, etc.) are either simplified to the time-delay equations, other variants of kinetic equations or treated by mean-field approaches (Sedhigi and Sengupta, 2007 [74]). However, various experimental techniques exist to study these effects such as Hi-C-Seq, Methyl-C-Seq, etc. and the network inference algorithms can be tuned to accommodate these parameters.

As mentioned, reconstructing the state-space of cancer networks and mapping cancer stem cell decision-making are intractable problems. The attractors of the Boolean networks may characterize the distinct cellular phenotypes and functional states of the developmental landscape (Kauffman, 1969 [62]; Huang, 2001 [75]). However, finding fixed-point attractors in Boolean networks is an NP-complete decision problem (Akutsu et al., 1998 [76]; Milano and Roli, 2000 [77]). To solve this, local search algorithms and optimization heuristics are employed (Milano and Roli, 2000 [77]). Currently, only the dynamics of small communities of a cancer network (subset of nodes) are studied to limit the size of the problem since exact solutions to large problems are considered unattainable (NP-hard). For example, finding a sub-graph within a larger graph network is an NP-hard problem (Barillot et al., 2013 [78]). Finding differentially mutated subnetworks of a larger gene-gene interaction network is likewise an NP-hard problem (Lu et al., 2016 [79]; Hajkarim et al., 2019 [80]). With chaotic attractors, the computational complexity and intractability is much greater. Finding attractors exhibiting chaotic behavior is an NP-hard problem (Pollack, 1991 [81]). In fact, finding an equation that describes how a system changes over time itself is an NP-hard problem (Cubitt et al., 2012 [82]). Therefore, finding the clique or master GRNs (Gene Regulatory Networks) controlling cancer stemness is an NP-complete problem.

**NONLINEAR DYNAMICS, FRACTALS AND CHAOS**

To encapsulate the complexity of tumor networks, cancer must be visualized within the framework of nonlinear dynamics. A review of nonlinearity in tumor biology is briefly painted in this section to illustrate how chaos and multifractality are universal characteristics of tumors and their complex dynamical networks (Ahmed, 1993 [83]).

Certain gene expression programs, as discussed above can dynamically switch a cell fate from a basin of attraction into another. Examples include the EMT/MET (epithelial-mesenchymal transition) switches. Sudden bursts and flickering of certain genes, transcription factors or morphogens (e.g., Wnt, Shh, Notch, etc.) can result in critical phase-transitions. The assumptions defined above demonstrate that current gene expression models assume the attractors formed by cancer networks' state-space to be fixed-point attractors or stable equilibria such as periodic orbits. However, cancer cell fates are complex, unstable attractors exhibiting spatiotemporal heterogeneity (Huang and Kauffman, 2013 [84]; Zhou et al., 2014 [85]; Li et al., 2016 [86]).

In principle, even minimal gene circuits can exhibit chaotic behaviors (Stokić et al., 2008 [87]; Hanel et al., 2010 [88]; Heltberg et al., 2019 [89]). For example, chaotic motifs (subgraphs within a complex network) were detected in few-node autonomous GRNs modeled by strongly coupled ODEs (Zhang et al., 2012 [90]). The study concluded that chaos can only appear in gene expression dynamics through competitions among different oscillatory modes of a GRN. When a single transcription factor, NF-kB, a well-described transcriptional regulator in cancer networks was modelled as a periodically forced nonlinear oscillator, chaotic dynamics emerged beyond a critical amplitude of TNF (Tumor Necrosis Factor) (Heltberg et al., 2019 [89]). Recently, these findings were confirmed with the Chaos Decision Tree algorithm, demonstrating the applicability of machine learning in detecting chaotic gene expression signals (Toker et al., 2020 [91]). Furthermore, chaotic behaviors are often equated with aperiodicity or disordered patterns of gene expression. For example, Nanog heterogeneity arises from fluctuations in gene networks and sudden burst-like puffs (intermittency) of transcription in the coexisting states (Smith et al., 2017 [92]). Although numerical/computational models of GRNs can exhibit chaotic behavior, the lack of time-series datasets remains a fundamental roadblock in experimentally detecting chaotic gene expression dynamics.

Takens (1980) [93] discussed fractal dimension analysis and entropy as algorithms for detecting the presence of strange attractors in dynamical systems. Electric cell impedance recordings were performed in rat's prostate cancers. The time-series, Fourier analysis of cancer micro-motions were assessed by Takens' theorem (time-delay coordinate embedding) to detect patterns distinguishable from a random signal. The attractor reconstruction showed positive Lyapunov exponents in the phase portraits (i.e., signature of chaos) (Posadas et al., 1996 [94]). Moreover, Posadas et al. (1996 [94]) demonstrated the Verhulst logistic model can simplify the period-doubling bifurcations observed in the cultured rat prostate cancer cells. However, attractor reconstruction techniques such as Takens's embedding theorem, convergent cross mapping and the calculation of Lyapunov exponents can become computationally difficult in higher-dimensional datasets such as single-cell RNA-Seq data.

Turing (1952 [95]) was the first to define morphogenesis as a reaction-diffusion system using a set of nonlinear partial-differential equations. A general reaction-diffusion system for two chemical species given in terms of their concentration densities u and v, is expressed as:

$$\frac{\partial u}{\partial t} = F(u,v) - du + D_u \Delta u$$

$$\frac{\partial v}{\partial t} = G(u,v) - dv + D_v \Delta v.$$

where the rate of change of u and v is the result of the production− (degradation + diffusion) terms (i.e., F and G describe the local production, d denotes the degradation rate, D denotes the diffusion coefficients and t is time) (Turing, 1952 [95]). Reaction-diffusion models characterize the growth-invasion dynamics of cancer ecosystems (Gatenby and Gawlinki, 1996 [96]; Ramis-Conde et al., 2008 [97]). The use of reaction-diffusion systems to predict cancer dynamics has been compared to forecasting weather patterns (Tang et al., 2014 [98]; Yankeelov et al., 2015 [99]).

Ivancevic et al. (2008) [100] showed a Lorenz-like chaotic attractor best describes the reaction-diffusion of cancer cells. Chaotic attractors emerge in the phase-space dynamics of tumor reaction-diffusion systems depending on changes in the order/control parameters, which vary amidst different phenotypes of a heterogeneous ecosystem (e.g., Diffusion coefficients, oxygen concentration, glucose level, tumor volume, diffusion from surface and growth-parameters) (Itik and Banks, 2010 [101]). While the Lorenz attractor has a fractal dimension of ~2.06, the cancer models showed a fractal dimension of ~2.03 indicating a chaotic attractor with Shilnikov-bifurcations (Itik and Banks, 2010 [101]). In a similar mathematical model, cancer was studied as a three-body ecosystem consisting of host, immune and tumor cells. Assuming initially logistic growth, the Lotka-Volterra (i.e., predator-prey) dynamics became chaotic as denoted by the period-doubling cascades in the bifurcation diagrams (Letellier et al., 2013 [102]; Khajanchi et al., 2018 [103]). The bifurcation analysis revealed Rössler-like strange attractors. Chaotic attractors in tumor pathology were determined to be prognostic indicators of tumor relapse and increased aggressiveness (Khajanchi et al., 2018 [103]). The emergence of Lorenz-like and Rössler-like attractors indicate cancer cell fates are *strange-attractors* of the Waddington landscape.

There is a striking similarity between pattern formation in reaction-diffusion systems and fluid convection models (Prigogine and Stengers, 1984 [104]; Ruelle, 1995 [105]). Reaction-diffusion even in simple chemical systems can give rise to turbulent patterns (Prigogine and Stengers, 1984 [104]; Ruelle, 1995 [105]). Lorenz (1963) [106] studied the Rayleigh-Benard convection system as a toy-model for weather turbulence, the approximate solutions of which became the Lorenz attractor. The critical Reynolds number required for Taylor-Couette and Rayleigh- Benard systems to transition to turbulence dynamics was shown to be in the order of ~100 which is relatively feasible at biologically relevant scales (Ruelle, 2012 [107]; 2014 [108]). While conventionally cells are assumed to be highly viscous structures with laminar protein flows, phase-transition to chemical turbulence can occur in reaction-diffusion systems as suggested by a recent theory based on Finite-elements Method simulation of bacterial Min protein systems (Halatek and Frey, 2018 [109]). The mammalian equivalent of Min proteins called PAR protein complexes are critical determinants of cancer stem cell bifurcation and tumor morphogenesis. The computational predictions were confirmed experimentally as well (Denk et al., 2018 [110]; Glock et al., 2019 [111]; Dang et al., 2019 [112]). The emergence of chemical turbulence at the onset of pattern formation (e.g., stem cell division and differentiation) are not accounted for in current cancer models. According to these findings, chemical turbulence at the onset of pattern formation is used as a synonym for spatio-temporal chaos, i.e., a broad distribution in the power spectrum and a low spatial correlation length reminiscent of Kolmogorov's energy spectrum. During chemical turbulence, both the amplitude and the phase of local concentration oscillations are strongly fluctuating, creating spiral waves of protein fluids followed by chaotic patterns.

The Navier-Stokes equations and its simplifications such as Darcy's law have been used to model tumor growth where tumor cells were considered as incompressible fluids (Yan et al., 2017 [113]; Vauchelet and Zatorska, 2017 [114]). Variants of the Navier-Stokes equations have successfully captured emergent behaviors such as active turbulence in cells (Marchetti et al., 2013 [115]; Bate et al., 2019 [116]). In active turbulence, turbulent flows are observed in cytoskeletal protein fluids and protein-mediated patterning within cells at relatively low Reynolds numbers (James et al., 2018 [117]; Martinez-Pratt et al., 2019 [118]). Turbulence dynamics have also been proposed to explain the Kolmogorov's power law decay observed in the frequency spectra of computationally simulated protein folding models. One of the studied proteins, Src kinases, are crucial drivers of cancer metastases and focal adhesion dynamics with the ECM (extracellular matrix) (Kalgin and Chekmarev, 2011 [119]; Andryuschenko and Chekmarev, 2016 [120]; Chekmarev, 2018 [121]). The Src kinases and associated adhesion proteins have shown capability of reprogramming cancer cells to benignity by restoring a subset of miRNAs to normal levels (Kourtidis et al., 2015 [122]). Conceivably, further investigation of the role of turbulence dynamics in the folding of cancer-associated proteins and intracellular chemical pattern formation may better direct the cancer reprogramming problem.

Fractal structures are ubiquitous in nature: think of the lungs, blood vessel networks, fluid turbulence and even heartbeats. Fractals show how complex structures can emerge from very simple rules which have no analytical form. Cancers are multi-fractal structures. Fractal image analysis and percolation networks show that in hypoxic conditions, tumor vasculature responds by growing into an extended fractal network known as angiogenesis through a heterogeneous ECM (i.e., invasion percolation) (Coffey, 1998 [123]; Baish and Jain, 2000 [124]). An emergent phenomenon known as vasculogenic mimicry, in which cancer cells under nutrient deprivation can spontaneously form blood vessel-like branching networks requires further investigation with regards to multifractality, cluster finding algorithms and percolation (e.g., Newman-Ziff algorithm, Leath algorithm for determining critical exponents, etc.). The fractal dimension can detect subtle changes in images and could potentially provide clinically useful information relating to tumour type, stage, and response to therapy (Lennon et al., 2015 [125]; Brú et al., 2008 [126]). Fractal characteristics of chromatin structure have been suggested as a diagnostic indicator of tumor self-organization. In general, the fractal dimension (FD) increases in tumor pathology (Metze et al., 2019 [127]).

The tumor texture has been modeled as a multi-fractal process, multi-fractional Brownian motion. A multi-fractal brain tumor segmentation was achieved using the AdaBoost classifier algorithm followed by machine learning algorithms such as decision trees, neural networks, or Support Vector Machine (SVM) as tumor sub-component classifiers. Islam et al. considered Diverse AdaBoost SVM algorithm followed by a Multiresolution Wavelet-Based FD Estimation for reconstructing tumor multi-fractality by predicting local scaling exponents in tumor architecture (Islam et al., 2015 [128]). Similarly, multi-scale wavelet analysis has been used as a measure of tumor multi-fractality in breast cancer diagnosis (Vasiljevic et al., 2015 [129]). Multifractal analysis was shown capable of distinguishing between histological images of the different chemotherapy responder groups as well. The f($\alpha$)max was identified as the most important predictive parameter, which represents the maximum of the multifractal spectrum f($\alpha$), where $\alpha$ is the Hölder's exponent (Vasiljevic et al., 2015 [129]).

**MACHINE LEARNING**

Artificial Intelligence and machine learning are at the frontier of tackling the cancer stemness problem. It is paving the future of precision oncology. The algorithms are trained to identify characteristic features in complex datasets and predict outcomes based on learned pattern recognition and signal processing. Deep Learning Networks are a subset of machine learning algorithms at the forefront of AI (Artificial Intelligence)-assisted clinical decision-making and *computational oncology* (Esteva et al., 2019 [130]; Topol, 2019 [131]). IBM Watson's individualized cancer diagnosis is a good example. Digital pathology, cell-lineage reconstruction from biopsies, drug screening, personalized pharmacogenomics, and identification of therapeutic targets/biomarkers are merely few examples of AI-assisted healthcare (Zhang et al., 2019 [132]; Silverbush et al., 2019 [133]; Huang et al., 2016 [134]; Ali and Aittokallio, 2019 [135]).

Convolutional Neural Networks (CNN) use multi-layered information processing and back-propagation to classify-predict complex signals such as images, speech and video data (LeCun et al., 2015 [136]). They are emerging as state-of-the-art approaches in carcinoma detection, classification and image segmentation analysis (Shan et al., 2019 [137]; Stower, 2020 [138]; Hollon et al., 2020 [139]). For example, DeepMACT is a Deep Learning-based pipeline for automated quantification of cancer metastases and therapeutic antibody targeting (Pan et al., 2019 [140]). Moreover, these networks can assess complex drug interactions in patients and are used in quality assessment of protein folding (Tong and Altman, 2009 [141]). For e.g., Deep learning nets cluster-classified the diverse subgroups in PDAC (pancreatic ductal adenocarcinoma), a highly heterogeneous and morbid disease (Zhao et al., 2018 [142]). Six clinically distinct molecular subtypes of PDAC were identified with 160 subtype-specific markers (Zhao et al., 2018 [142]). Deep learning can predict microsatellite instabilities in gastric cancers and thus, improve immunotherapy decisions in patients (Kather et al., 2019 [143]). Multi-omics data and machine learning are used to predict metabolic pathways in cancer resistance (e.g., signaling networks associated to hypoxia, aerobic glycolysis/Warburg effect, etc.) (Castello and Martin, 2018 [144]).

As mentioned, recent findings show cells expressing CSC-associated cell membrane markers in Glioblastoma (GBM) do not represent a clonal entity defined by distinct functional properties and transcriptomic profiles, but rather denotes a 'plastic state' that most cancer cells can adopt (Dirkse et al., 2019 [145]; Neftel et al., 2019 [146]). The findings show that GBM cells exist in four main clusters corresponding to distinct neural cell types (Neftel et al., 2019 [146]). Basic machine learning algorithms were critical in paving these results. The t-SNE projections of GBM scRNA-Seq found cellular subpopulations resembling different expression subtypes co-occurring in the same tumor and adapting to heterogeneous phenotypes in disease progression (Patel et al., 2014 [147]; Yuan et al., 2018 [148]). Other findings suggest the conserved neural trilineage, a hierarchy with glial progenitor-like cells at the apex in GBM (Couturier et al., 2020 [149]; Pang et al., 2019 [150]). A droplet-based scRNA-Seq of patient-derived GBM cells, generated two or three distinct clusters by t-SNE and the Louvain community structure detection algorithm. Following, PCA (principal component analyses) was used to assess intra-tumoral heterogeneity within the enriched GSCs (glioma-stem cells) of each grouping. The data suggested that GSCs are organized into progenitor, neuronal, and astrocytic gene expression programs, resembling a developing brain. However, the findings do not discuss the implications of microenvironment-dependent phenotypic plasticity and interconvertibility of cell fates. The time-progression (transitions) in between these clusters are ambiguous given they are dynamical systems. The key point emphasized herein is the importance of machine learning algorithms in interpreting cancer data and stemness networks.

Liquid biopsies are enriched with circulating tumor cells (CTCs), circulating tumor DNA (ctDNA), tumor-derived extracellular vesicles, free nucleosomes and many other oncogenic biomarkers presenting a vast repertoire of information for tumor detection and prognosis. Exosomes are nanoscopic (~30-100 nm), heterogeneous packets of information released by cells forming long-range, intercellular communication networks. They are emergent, reprogramming machineries used by cancer ecosystems to regulate cell fate dynamics (Guo et al., 2019 [151]). That is, exosomes can reprogram distant tissue microenvironments into pre-metastatic niches and horizontally transfer malignant traits such as therapy resistance to promote aggressive cancer phenotypes (Steinbichler et al., 2019 [152]; Keklikoglou et al., 2019 [153]). For example, exosomes derived from an hESC (human embryonic stem cell) microenvironment has been shown to suppress cancer phenotypes and reprogram a subset of malignant phenotypes to healthy-like, benign states (Camussi et al., 2011 [154]; Zhou et al., 2017 [155]). An elevated expression of the Yamanaka factors and plausibly microRNAs in the nano-vesicles were identified as the causal mechanism for the cell fate reprogramming. Machine learning is reshaping our understanding of the cybernetics of these nanoscale communication networks.

Machine learning algorithms applied on Intraoperative Raman spectroscopy can distinguish brain cancer cells from normal brain tissue (including both invasive and dense cancers) with an accuracy, sensitivity, and specificity > 90% (Jermyn et al., 2015 [156]). Background subtraction algorithms (e.g., the rolling ball algorithm), feature extraction and autofluorescence removal algorithms were employed to distinguish the cancer cells' Raman bands from those of healthy tissues (Brusatori et al., 2017 [157]; Zhao et al., 2007 [158]). Furthermore, machine learning is paving minimally invasive, cancer screening with liquid-biopsy derived exosomes characterization. Simple machine learning algorithms such as binomial classifiers (GLMnet) with the caret R package, can be trained to classify tumor-associated methylome patterns in cell-free plasma DNA (Shen et al., 2018 [159]). Therefore, machine learning of aberrant epigenetic signatures in ctDNA are diagnostic indicators for cancer patients.

Moreover, patient-derived exosomes were analyzed using Raman spectroscopy with a 785 nm laser at ~5mW irradiance onto samples plated on calcium fluoride substrates. Cluster analysis and interpolation found that stem cell-derived exosomes grouped differently when subjected to simple dimensionality reduction techniques like PCA (Gualerzi et al., 2019 [160]). Similarly, surface enhanced Raman scattering (SERS) signals of exosomes from normal and NSCLC (Non-small-cell lung carcinoma) cells on Gold-nanoparticle substrates were performed. The Raman spectra of cancerous exosomes showed unique peaks in the vibrational bands distinguishing NSCLC exosomes from healthy clusters when subjected to spectral decomposition by Principal Component Analysis (PCA) (Shin et al., 2018 [161]; Rojalin et al., 2019 [162]). Correlation analysis further revealed potential exosome surface protein markers, where cancerous exosomes had similar spectra (Park et al., 2017 [163]; Shin et al., 2018 [161]).

Similar findings with peak fitting algorithms and the MCR-ALS algorithm (multivariate curve resolution-alternating least squares) on Raman spectra were used to cluster-classify pancreatic cancer exosomes on gold-nanoparticle plated SERS substrates (Banaei et al., 2017 [164]). Another recent study shows neural networks optimized with machine learning algorithms such as principal component analysis-linear discriminant analysis (PCA-LDA) and binary classifiers like support vector machines (SVM) can distinguish oral cancer patients from healthy individuals by characteristic signatures in the Fourier transform infrared (FT-IR) spectroscopy of their salivary exosomes (Zlotogorski-Hurvitz et al., 2019 [165]).

In a recent finding, Random Forest classification of exosomes using mass spectrometry-based proteomic profiling revealed cancer detection specificities and sensitivities of 90% and 94% for tissues, and 95% and 90% for plasma, respectively (Hoshino et al., 2020 [166]). Plasma- derived exosome cargo and cell surface markers were shown capable of distinguishing among cancer types in patients (Hoshino et al., 2020 [166]). These findings collectively suggest artificial intelligence and machine learning algorithms can pave the early-stage detection, prognostic screening, and classification of cancers from patient-derived liquid biopsies.

**NETWORKS RECONSTRUCTION**

Currently, both statistical and machine learning algorithms are used in reconstructing cancer GRNs and PPI (protein-protein interaction) networks under a probabilistic Boolean structure (Kar et al., 2009 [167]; AlQuraishi et al., 2014 [168]; Fumia et al., 2013 [169]; Li et al., 2020 [170]). PPI networks are widely used to study cancer network dynamics where given the changes in protein concentration in time, the factors most connected in the regulatory networks are determined. Various algorithms and software packages utilizing information-theoretic measures and community detection algorithms are used to for computing the shortest paths on the protein interactome between differentially expressed genes in transcriptomic datasets (Zhang et al., 2017 [171]; Liu et al., 2017 [172]). Machine learning algorithms can also be used to infer mutated genes in a network (Vandin et al., 2011 [173]). As an example, diffusion processes and Monte Carlo approaches have been used to define a local neighborhood of influence for each mutated gene in datasets from glioblastoma and lung adenocarcinoma samples (Vandin et al., 2011 [173]).

Various information-graph theoretic methods are employed to reverse engineer cancer networks from multi-omics datasets, providing key insights into the molecular underpinnings of cancer stemness (Zhang et al., 2017 [174]; Mallik and Zhao, 2020 [175]). For example, Cytoscape, an open source platform for visualizing molecular gene-interaction networks revealed candidate barrier genes in the RNA-Seq data of MCF7 breast cancer cells preventing their iPSC reprogramming (Bang et al., 2019 [176]). Cytoscape uses Gene set enrichment analysis (GSEA), a method of statistical approaches that assigns a score to potential gene-gene interactions by identifying significantly enriched or under-expressed modules/clusters of genes. PRMT6, MXD1, and EZH1 were highly expressed in cancer cells during reprogramming regardless of the ectopic expression of OSKM (i.e., acronym for the Yamanaka factors), indicating that these genes could be obstacles in the early stage of MCF7 cell reprogramming (Bang et al., 2019 [176]). However, the pathways and networks are represented as static processes in Cytoscape (Merico et al., 2009 [177]). More detailed mathematical representations exist for handling network dynamics and kinetic models (e.g., VirtualCell, E-cell, etc.). Some examples of these network visualization algorithms with information flow inference are discussed below. To better understand the cancer stemness problem and cancer reprogramming, the mathematical framework of these graph-theoretic networks must be first appreciated.

Bayesian Network inference algorithms are well-suited for inferring GRN topologies from measured gene expression levels due to their ability to reverse engineer the cause-effect relationships in a graph structure (Mehreen and Aittokallio, 2019 [178]; Jausen et al., 2003 [179]). Although Bayesian networks can capture causal relationships, learning such networks from a given dataset is NP-hard (Luo et al., 2020 [180]). A Bayesian network is a type of a Probabilistic Graphical Model (PGM) where each node represents a random variable $G_i$ corresponding to the expression level of gene i. The dependencies are directed and acyclic. The continuous gene expression levels are discretized into three bins: under-expressed is denoted -1, control as 0 and over-expressed as 1. An edge going from $G_i$ to $G_j$ indicates the gene i (transcription factor) regulates gene j in this network. Then a local probability model describes the conditional probability for the expression level of a gene, given the expression levels of its neighboring genes. Dynamic Bayesian Networks (DBN), an extension of the Bayesian network considers temporal information in the data. The gene expression microarray or RNA-Seq data is sliced into individual time points, and each slice becomes a random variable in the DBN network.

Bayesian models are also used in sorting cell phenotypes from single-cell transcriptomic data (Shun et al., 2019 [181]). As stated, although the Bayesian network is a prominent method to somewhat infer gene regulatory networks, learning the Bayesian network structure is an NP-hard problem (Ahmad et al., 2012 [182]). Once the general network topology is deduced from these methods, the GRN modules are interwoven using distance metrics and clustering algorithms (Pajtler et al., 2015 [183]; Gladilin, 2017 [184]). Clustering methods identify the distinct cell types in a population of cells' gene expression profiling (Creixell et al., 2012 [185]). Statistical measures such as correlation scores, sensitivity, ROC (Receiver Operating Characteristic) curves, etc. are further assessed to validate the statistical reliability of these networks with regards to the presented data. Correlation analysis validates whether two genes expressed in the given condition are co-expressed. However, it does not infer causal relationships and dynamics. Other follow the crowd approaches to networks reconstruction include regression networks and Mutual Information (MI). Mutual Information is an information-theoretic algorithm that detects nonlinear relationships in gene expression when the correlation seems to be inadequate. It measures how much information one random variable tells us about another. For example, MI identified regulatory modules of the Myc transcription factor, a key regulator of cancer stemness (Basso et al., 2005 [186]). DREMI (Density Resamples Estimate of Mutual Information), a computational algorithm based on MI, can map the information flow and developmental trajectories in complex single-cell datasets with drug perturbation analysis (Krishnaswamy et al., 2018 [187]). DREMI well-predicted the time-varying information flow in the GRNs of breast cancer cells undergoing EMT (epithelial-mesenchymal transition) (Krishnaswamy et al., 2018 [187]).

The general approach to scRNA-Seq-based cell lineage clustering and pseudotime inference is based on pre-processing steps (feature extraction) followed by a combination of machine learning algorithms to infer the information flow. The raw data reads are filtered and selected for most differentially expressed genes in the cells where false counts remain a problem (Mohammadi et al., 2019 [188]). The data is then fed into the machine learning algorithms as a gene expression matrix of cells (columns) by genes (rows), followed by dimensionality reduction techniques like PCA (principal component analysis) or t-SNE, etc. Then, network-graph theory approaches reconstruct the spatial neighborhood of cells with statistical inference. For example, K-means clustering, or K-Nearest Neighbor (KNN) graphs are used in this task as supervised and unsupervised techniques, respectively. Next, optimization algorithms are used to find cell fate trajectories and regulatory modules. Community detection algorithms or partial-information decomposition algorithms assess the distinct cell clusters with similar gene expression profiles. Various dispersion cell-cell variability (pairwise) measures such as covariance, correlation, Bayesian inference and metric entropy are used to better distinguish cell clusters and reconstruct a distance/similarity matrix (Teschendorff et al., 2017 [189]). A typical workflow for network visualization that incorporates these statistical approaches is Weighted Gene Co-expression Network Analysis (WGNCA) followed by Gene Set Enrichment Analysis (GSEA). The gene expression profile is grouped into bins from an available Gene Ontology (GO) followed by statistical approaches such as enrichment scores (e.g., Kolmogorov-Smirnov-like statistic) and statistical significance testing to calculate the association between differentially expressed gene clusters.

Broadly speaking, there are two categories of methods widely used for reconstructing developmental trajectories from single-cell multi-omics. The first, often referred to as pseudotime ordering methods rely on dimensionality reduction followed by an ordering step. The second learns a probabilistic branching model to represent the developmental process (Fischer et al., 2019 [190]). Given initial cell assignments, the branching probabilities can be inferred using standard Maximum Likelihood Estimation (Lin and Bar-Joseph, 2019 [191]). However, the ordering is based on a very limited set of values for each cell and the nonlinear dynamics of cell fates are neglected. As stated, machine learning algorithms are the current state-of-the-art approaches in processing the information flow from single-cell data to infer signaling networks (Saelens et al., 2019 [192]; Yang et al., 2019 [193]). They are used for identifying driver mutations in single-cell cancer datasets and predicting molecular subtypes (Chan et al., 2017 [194]; Gan et al., 2018 [195]; Ozturk et al., 2018 [196]). The following are some examples of network trajectory inference and visualization algorithms in practice suitable for time-series gene expression datasets.

Topslam estimates state-transition pseudotime by mapping individual cells to the surface of a Waddington-like landscape with Bayesian Gaussian Process Latent Variable Model (GPLVM), a nonlinear probabilistic dimensionality reduction (Zwiessele and Lawrence, 2017 [197]). These methods project high dimensional data into 2 or 3 components, whereby distances are interpreted as cell-cell variability. Seurat is another pipeline that can be used as a pre-processing tool for cancer data sets. Seurat is a scRNA-Seq correlation and clustering computational tool (Butler et al., 2018 [198]; Stuart et al., 2019 [199]). The identified clusters can be further sorted by other algorithmic packages into flow networks. Many of the cancer stem cell findings discussed above used the Seurat algorithm. The Python-based implementation ScanPy, very much alike Seurat, is an easier platform to operate despite less-available algorithms (Wolf et al., 2018 [200]). While Seurat and Scanpy provide differentiation mapping of scRNA-Seq data using dimensionality reduction techniques like PCA, t-SNE, and UMAP (Uniform Manifold Approximation and Projection), a higher-resolution visualization is provided by PHATE (Potential of Heat-diffusion for Affinity-based Transition Embedding) (Moon et al., 2019 [201]). PHATE is an unsupervised low-dimensional embedding algorithm used in visualizing emergent, nonlinear patterns and structures in static (i.e., snapshots of) single-cell gene expression datasets (Moon et al., 2019 [201]).

The Stochastic Simulation Algorithm (SSA), also known as the Gillespie algorithm provides make use of Monte Carlo type simulations to approximate solutions to the chemical Master equation (i.e., Fokker-Planck equations) and reconstruct a stochastic multidimensional GRN. Although generally applicable, such algorithms are computationally expensive. The SELANSI (Semi-LAgrangian Simulation) algorithm is a software toolbox that uses a semi-Lagrangian method to solve partial integral differential equations such as the Fokker-Planck equations (Pájaro et al., 2017 [202]). Due to these equations in its framework, the algorithm can accommodate sudden abrupt changes in the kinetics of transcriptional regulation such as protein bursting, which occurs whenever mRNA degrades faster than protein products (Pájaro et al., 2017 [202]).

sgnesR (Stochastic Gene Network Expression Simulator in R) is an R package that simulates GRNs using the Gillespie algorithm (Tripathi et al., 2017 [203]). The sgnesR package provides a GRN structure for time-series gene expression data. The package allows the user to fine-tune various parameters to model the GRN dynamics and optimize the bias-variance trade-off (i.e., hyperparameter tuning). The initial population parameter assigns the initial values of promoters, RNAs and proteins for all the differentially expressed genes in the network. Binding and unbinding rates are assigned to the transcription kinetics based on experimental measurements. Delay parameters can assign a delay-time for RNAs, proteins in synthesis (translation) and degradation reactions, and to the timing of transcription reactions (Tripathi et al., 2017 [203]).

CellRouter is a graph-theoretic, flow network-based trajectory detection algorithm (Lummertz da Rocha et al. 2018 [204]). First, dimensionality reduction (PCA, t-SNE, Diffusion map, etc.), is performed on the single-cell dataset. Following, a KNN graph is assessed. Then, the Jaccard index finds the similarity between two cells where if they belong to the same cluster, a high correlation is observed. The Louvain community structure detection algorithm configures the populations by assessing weights of the graph by similarity of cell-cell interactions. As such, a source-to-sink directed graph is projected mapping the GRN flow network where the nodes are connected by flow arrows with the weighting indicating their capacity. The trajectories are found with the Bellman-Ford algorithm, a type of cost flow optimization algorithm and ranked by total flow, cost and length between vertices. The Bellman-Ford algorithm computes the shortest paths from a single source vertex to all vertices in a weighted graph. It is slower than the Dijkstra's algorithm but can handle graphs with negative edge weights. Lastly, the optimized trajectories are ranked by GRN scores given by Pearson-Spearman correlation. Corresponding heat maps and dynamic curves of regulators are further projected to obtain trajectories of the identified clusters, displayed as a GRN flow network.

Single cell Energy path (scEpath) is a method for mapping the energy landscape of single-cell gene expression datasets. It reconstructs cell fate trajectories with information-theoretic measures such as maximum entropy (Jin et al., 2018 [205]). The scRNA-seq data is initially pre-processed and filtered by removing low gene expressions. By calculating the Spearman correlation of the adjacency matrix, scEpath builds a GRN, calculates the normalized energy between expressions, and performs a linear dimensionality reduction with PCA. Data then undergoes cell clustering using an unsupervised schema called single-cell interpretation via multikernel learning (SIMLR). From the cell clusters, statistical distributions are used to find the transition probabilities. To infer cell lineages, scEpath first constructs a probabilistic directed graph with maximum probability flow, equivalent to finding a minimum directed spanning tree (MDST) from Edmonds' algorithm. Finally, scEpath uses the R "princurve" package to fit a principal curve of the clusters to compute the pseudo-times. Alike scEpath, numerous entropy-based algorithms are emerging for quantifying single-cell potency in the Waddington landscape (Shi et al., 2020 [206]).

The pre-processing in scEpath is similar to Slingshot, a trajectory cell-lineage classification algorithm in which diffusion maps, PCA, etc. are used as dimensionality reduction followed by model-based clustering (Street et al., 2018 [207]). The expectation-maximization algorithm, a class of maximum likelihood measures, and Bayesian inference are then applied to the dataset. As an alternate, k-means clustering can be assessed to reconstruct relative cell fate commitments. Lastly, minimum spanning trees (MST) are performed on the clusters using Prim's algorithm to infer cell fate branching. Some of the best machine learning approaches in mapping cell fate choices consist of t-SNE followed by MST on single-cell datasets. However, neural networks are generally preferred for larger datasets.

Markov random fields/Markov networks are also emerging as powerful algorithms in reconstructing cancer networks. For example, ssNPA (single sample Network Perturbation Assessment) is a new network-based approach which learns a causal graph from single-cell gene expression data and builds a Markov blanket-based predictive model to infer the GRN (Buschur et al., 2020 [208]). ssNPA identified molecular subtypes in lung and breast cancer scRNA-Seq with a better accuracy than many alternative network inference algorithms (Buschur et al., 2020 [208]).

The Hopfield network is a recurrent neural network displaying associative memory. It is a breakthrough in the field of cybernetics and employed as a searching heuristic in classification problems, optimization problems and pattern recognition (Hopfield, 1982 [209]; Hopfield and Tank, 1985 [210]). The Ising model, the statistical fabric of Hopfield neural networks, demonstrates a problem faced in current computational physics namely that not all systems have closed form solutions (i.e., NP-completeness). The Ising model is closely related to an open question in complexity theory which concerns whether sometimes the only way to find a solution is to study all possibilities by exhaustive searching (Bossomaier and Green, 2000 [211]; Szedlak et al., 2014 [212]). Lang et al. (2014) [213] using single-cell data demonstrated the Hopfield neural network can reconstruct the distinct cell states of the Waddington landscape as spin glass topologies. A subspace representative of attractors is obtained by taking the orthogonal projection of key transcription factors' gene expression. Correlation measures and statistical significance testing methods (e.g., z-scores) then defined the distinct clusters corresponding to the gene regulatory networks (Lang et al., 2014 [213]). Partially reprogrammed cell fates emerged as hybrids co-expressing signals from multiple cell fates in the landscape (i.e., spurious attractors) (Lang et al., 2014 [213]).

Hopland is a continuous Hopfield neural network-based algorithm which interprets single-cell gene expression data for Waddington landscape reconstruction (Guo and Zheng, 2017 [214]). The cell fate attractors are reconstructed based on gene to gene expression correlation from the gene-expression matrix. First, Isomap dimensionality reduction is performed. Isomap generally consists of a KNN graph followed by geodesic calculations between the identified cell clusters from the Dijkstra's algorithm and multidimensional scaling. Each gene is modelled as a neuron in the network, the cumulative energy values of which add up to each cell fate on the landscape. The fast-marching algorithm devised for solving the Eikonal equation was used to calculate the geodesic distances on the landscape as the weights of edges connecting the cells on a triangulated-mesh grid scheme. A Gaussian-mixture optimization algorithm was used to infer the parameters and the mean values of the outputs (gene expression values) from the identified cell clusters. Then, the Gradient descent algorithm optimizes the Hopfield network whereby the activation values (weights of the edges on the GRN graph) undergo a relaxation process by minimizing the Lyapunov energy function. The energy minimization derives the local attractors of the network state space. The Lyapunov energy function gives the energy values corresponding to the cell states, where a high energy corresponds to less-differentiated states (hills), while low energy indicates differentiated cell fates (apexes and valleys). The Lyapunov energy function is given by:

$$E = -\frac{1}{2}\sum_{i=1}^{N}\sum_{j=1}^{N} W_{ij} U_i U_j + \sum_{i=1}^{N} I_i U_i + \sum_{i=1}^{N} \delta_i \int_{0}^{u_i} g^{-1}(u) du$$

here g is the activation function, W is the weights assessed from gene expression data as described above, $\delta_i$ is the gene signal degradation rate, and $I_i$ is the combination of propagation delays and regulations from the environment classified as 'noise', and $U_j = g_j(V_j)$, analogues of magnetic dipole moments in the Ising model corresponding to the nearest neighbor energies of neurons (genes), where V is the neuron outputs for N genes. To simulate dynamic trajectories, the 1st order Runge-Kutta finite difference method is used to solve the updating of neurons (genes) as an ordinary differential equation while the Euler-Maruyama method is used for solving stochastic differential equations such as the Langevin/Fokker-Planck equations or any appropriate Master equation underlying the network dynamics. These two techniques are applied for high molecular concentrations inferred by the gene-gene correlations, whereas the Gillespie algorithm is used by the neural network for low copy numbers (i.e., stochastic fluctuations) (Guo and Zheng, 2017 [214]). To visualize the landscape topography, the GPLVM dimensionality reduction technique is applied and lastly the MST (Minimum Spanning Tree) algorithm identifies the minimum trees connecting the cell type clusters. Hyperparameter tuning allows the optimal learning rate and regularization as well. While the Hopland algorithm remains unexplored in cancer datasets, a Hopfield network was recently shown efficient in revealing cancer attractors related to molecular subtypes identified in the bulk RNA-Seq data of paired breast tumor and control samples from 70 patients (Conforte et al., 2020 [215]). A modest correlation was revealed in this simple tumor case, for which there is already abundant evidence of drug efficacy (Conforte et al., 2020 [215]). The next step would be to apply these algorithms to more complex cancers such as GBM.

Furthermore, the application of a neural network-based classifier on whole-genome RNA-Seq data in cancer patients showed a near 86% success rate in diagnosing complex, metastatic cancers (n= 201) and 99% success rate on primary cancers (Grewal et al., 2019 [216]). However, esophageal cancers and adenocarcinoma were often misclassified. Neural networks were trained and optimized for the highest probabilistic inference of pathological assessment. A variety of machine learning algorithms including ANN (Artificial Neural Networks), Bayesian networks, multi-class Support Vector Machines and decision trees were used as predictive models for the cross-validation and training in assessing pathological staging. Then, confidence scores were assigned in their performance evolution. The weight analysis of the neural network was used to identify the genes most important in class prediction of tumor subtype (Kourou et al., 2015 [217]; Grewal et al., 2019 [216]).

Deep Learning neural networks are emerging in biological networks reconstruction and deciphering differential gene expression patterns (Tian et al., 2019 [218]; Tasaki et al., 2020 [219]). For instance, scDeepCluster uses a model-based cluster analysis through multi-layered neural networks. As discussed in the previous examples, multi-kernel spectral clustering methods and community detection methods are amidst the most used unsupervised learning methods in cell lineage reconstruction. However, such methods do not account for the false zero (low RNA capture) counts and scalability (dimensionality reduction) issues in scRNA-Seq cluster analysis. Deep learning embedded clustering was proposed as a solution (Tian et al., 2019 [218]). The autoencoder, a type of deep neural networks was used to replace the mean square error (MSE) loss function with a zero-inflated negative binomial (ZINB) model-based loss function. The autoencoder performs a nonlinear function mapping of the scRNA-Seq read count matrix to a low dimensional latent space. Following, cluster analysis is performed by the Kullback–Leibler (KL) divergence which characterizes the relative information entropy in clusters. The KL divergence was followed by the 'deep embedded clustering' (DEC) algorithm along with noise-reduction techniques to classify the distinct cell fate clusters (Tian et al., 2019 [218]). Other less popular Deep Learning methods such as imputation algorithms are available as well. For instance, SAUCIE (Sparse Autoencoder for Unsupervised Clustering, Imputation, and Embedding) is a Deep Neural Network capable of single-cell data clustering, visualization, and denoising/imputation in longitudinal datasets (Amodio et al., 2019 [220]).

The above discussed are only a sample of the many algorithmic packages and tools from complex systems theory available for cell fate trajectory analysis and cancer networks reconstruction. For example, evolutionary algorithms have also been used to infer topologies of Gene Regulatory Networks (Aguilar-Hidalgo et al., 2015 [221]). Some chaotic attractor reconstruction methods have also been attempted with limited success such as convergent cross-mapping, used to identify causal trajectories in the transcriptional modules of hematopoietic cytokine networks (Krieger et al., 2018 [222]). The efficiency of the above-discussed algorithms is model-dependent. To illustrate, Topslam outperformed Hopland in dissecting the scRNA-Seq of mouse embryo development using GPLVM and MST alone. Furthermore, the datasets in these algorithms were highly filtered where low-gene expressions are removed from analysis. Although the law of mass action may be held as a reasonable justification, with time-series analysis a sudden burst in the filtered-out signal can have cascading effects on the gene network dynamics; a signature of chaotic systems (i.e., sensitive dependence to perturbations). Then, how does one distinguish a random network from that of a chaotic network? How do we infer causal relationships in gene network dynamics? A branch of complexity theory known as Algorithmic Information Dynamics (AID), is emerging as a solution.

**ALGORITHMIC INFORMATION DYNAMICS**

As discussed, the study of large nonlinear systems of equations are dealt with approximation tools such as mean-field approaches, model reduction and machine learning. These methods have limited prospects in studying the nonlinear kinetic equations governing GRN networks. Detecting strange attractors in GRN reconstruction remains an intractable problem. However, AID provides an alternative approach to reconstruct dynamical systems with disordered/chaotic. Zenil et al. (2019) developed an algorithmic calculus, Algorithmic Information Dynamics (AID), to overcome these challenges in network science. It is an alternative to the above-discussed graph-theoretic methods of data science which primarily rely on statistical inference, machine learning and classical information theory – which often fail to account for causal structures in network information flow.

Algorithmic Information Dynamics (AID) provides computational tools derived from Algorithmic Information Theory (AIT) in combination with perturbation analysis to study dynamical systems in program space (Zenil et al., 2019 [223]; 2020 [224]). AID is emerging as a powerful tool for studying the cybernetics of complex networks wherein perturbation analysis is used to steer or reprogram a network towards a specific direction (e.g., cell fate commitment). The application of AID in biological networks reconstruction has been well-established by Zenil et al. (2019 [223]). The mere notion of 'reprogramming' indicates cells are computational systems whose fates correspond to specific chemical programs (Zenil et al., 2017 [225]). According to Zenil et al. (2017) [225], cancer can be defined as a computer program in an infinite loop with no halting condition. Unhealthy states of the dynamical immune system were defined as *strange attractors*, while healthy states were considered as fixed-point attractors in state space (Zenil et al., 2017 [225]). The term strange attractor was first coined by Ruelle and Takens to characterize the multifractal chaotic patterns observed in the phase-space bifurcations of fluid turbulence (Ruelle, 1995 [105]; Lanford, 1982 [226]).

K-complexity (i.e., Kolmogorov-Chaitin complexity) also referred to as algorithmic complexity, is a robust measure of complex network dynamics vastly unutilized in current cell-lineage reconstruction approaches. Kolmogorov complexity studies the length of the shortest computer program that represents a system and may characterize what constitutes a cause as opposed to randomness in dynamical systems. Although K-complexity shares many properties with Shannon's entropy, it provides a more meaningful measure of complexity than the latter. The algorithmic complexity of a string s is given by $K(s|e) = \min(|p|: U(p,e) = s)$, where p is the program running in a universal Turing machine U with input e. An object s is random (non-causal) if the algorithmic complexity K(s) of s is about the length of s itself (in bits) (Toscano et al., 2014 [227]; Zenil et al., 2016 [228]). Algorithmic complexity can be analogously seen as a measure of compressibility. However, the major drawback of K(s) as a complexity measure is its incomputability (Zenil et al., 2018 [229]).

Algorithmic probability considers the probability of a discrete object to be produced by a universal Turing machine. It captures network dynamics on the basis of two principles: Occam's razor (i.e., algorithmically the most probably outcome is the shortest (simplest) algorithmic description), and the principle of multiple expansion (i.e., to keep all explanations consistent with the data and Bayes' rule). The expected probability that a random program p running on the Turing machine produces the string s upon halting is given by the Solomonoff-Levin algorithmic probability measure: $P(s) = \sum_{p:T(p)=s} \frac{1}{2^{|p|}}$.

As this measure is incomputable, numerical approximation methods exist for indirectly approximating the K-complexity. Most attempts to approximate algorithmic complexity K(s) have been made by lossless compression algorithms like LZ (Lempel-Ziv) or LZW and MML/MDL which are essentially estimators of entropy rates. Zenil et al. introduced the Coding Theorem Method (CTM), which establishes an equality between the algorithmic probability P(s) and the Kolmogorov complexity K(s), formally expressed as:

$$P(s) = 2^{-K(s)} + c$$

where c is some constant, or equivalently, $K(s) \approx -log_2 P(s)$. The Algorithmic Coding Theorem Method states that an object of a short computer program is more likely to be generated at random than an object generated by a longer computer program (i.e., complex systems). While lossless compression algorithms can only deal with statistical regularities within a certain window of length size, CTM has a potential to identify non-statistical regularities but remains computationally expensive. Therefore, Zenil et al. (2019) [223] developed a toolbox of analytical techniques that facilitate the search for sub-network structures such as motifs, clusters and modules and their statistical feature extraction. Some of the algorithms that serve as approximation measures for algorithmic complexity include BDM, MILS and MARPA.

BDM (Block Decomposition Method) is a weighted version of Shannon's entropy that introduces algorithmic randomness. It extends the CTM to quantify algorithmic randomness by decomposing data into pieces small enough that an exhaustive search finds the set of all computable models able to generate each piece. A sequence of computer programs smaller than their matched patches constitutes a sufficient test for non-randomness (Zenil et al., 2019 [223]; 2020 [224]). Minimal Information Loss Sparsification (MILS) is a method to identify the neutral elements that have null or negligible algorithmic information content value in a network. That is, by deleting a certain number of graph edges, it identifies elements that can be safely removed with maximal preservation of graph-theoretic measures such as edge betweenness, clustering coefficients and node degree distribution of the network. The Maximal Algorithmic Randomness Preferential Attachment (MARPA) algorithm is alike the reverse algorithm to MILS which seeks to maximize the information content of a graph by adding new edges or nodes. It approximates a network of a given size that has the largest possible algorithmic randomness (i.e., Erdos-Rényi graph).

The applicability of these tools in biological networks reconstruction was demonstrated on the experimentally validated TF network of E. coli (Zenil et al., 2019 [223]). Classical measures such as Shannon entropy and lossless compression were shown to be less sensitive than BDM in encapsulating the complexity of the network's dynamics in program space. Small changes were not captured by the compression algorithms followed by enrichment analysis. These findings were further validated in T-cell differentiation and multiple human cell type datasets including HPSC and ESC stem cells, thereby demonstrating the applicability of AID in deciphering CSCs (cancer stem cells) networks as dynamical systems.

The typical AID flow map for attractor landscape reconstruction is as follows: a validated gene regulatory network with its associations and categorizations of gene-protein interactions is selected from a gene ontology database. Genes that protect the network's structural and functional integrity were assessed with algorithmic perturbation analysis (i.e., remove or knockout (*mutate*) a node or edge in the network). Following the information spectral analysis, clustering methods are used (e.g., k-means algorithm) to find the classified genes having the most negative or positive information values in the differentiation networks. Gene ontology (GO) and pathway enrichment analysis further classifies gene clusters in association to their specified functions by cross-validating with gene-databases and literature (text) mining. CellNet, a network science-based computational platform can reverse engineer the differentiation landscape where the activator-inhibitor dynamics of each gene is mapped as peaks and valleys of a Waddington epigenetic landscape. In principle, stem cell fates would be attractors fluctuating in between higher peaks of the epigenetic landscape (high entropy). The above-mentioned AID algorithms can quantify the changes in the shallowness (depth), length and the number of attractors on a biological network's state-space (Zenil et al., 2019 [223]; 2020 [224]). Therefore, AID is a promising algorithmic approach to steer for attractors in the Waddington landscape reconstruction of cancer networks. This sub-section is merely a scratch-the-surface to demonstrate the applications of AID in deciphering complex dynamical systems.

**ALGORITHMIC PROSPECTS**

While time-series datasets will model cancers as dynamical systems, the algorithms for their networks and developmental landscape reconstruction must be adapted to time-series analysis. As discussed, delay-coordinate embedding algorithms are limited to the size of a dataset. The following are prospects to further conceive the marriage of computational algorithms and time-series cancer datasets.

Domain translation methods (i.e., image-to-image translation) are simple embedding algorithms capable of attractor reconstruction. Given pairs of elements from two different domains, the methods consist of learning a mapping from one domain to another by linking these paired elements (Ayed et al., 2020 [230]). For instance, fluorescent reporters can help track the concentration changes of labelled proteins involved in tumor pattern formation and gene regulation with time-lapse imaging. The scRNA-Seq reads of these cells can be assessed in simultaneity. The domain translation method can then train a neural network for the residual mapping of these data between different time-points. Object edge detection and similar algorithms can track multiple arrays of fluorescent-labelled gene products and protein flows in time lapse imaging of cells. This is a counting problem. Image density maps can be predicted using supervised learning methods optimizing the loss based on the MESA-distance (Maximum Excess over SubArrays). Trajectories can then be optimally learnt by applying a Gaussian peak to label the centroid of cells and map the cell from frame to frame (Lemptisky and Zisserman, 2010 [231]). Regression networks with convolutional redundant counting are currently used to tackle counting problems in complex dynamical systems such as traffic flows, crowds, and even cells. Generic Matching Network (GMN) architectures can also accomplish the counting and potentially map strange attractors in tumor pattern formation systems (Cohen et al., 2017 [232]; Lu et al., 2018 [233]).

GANs (Generative Adversarial Networks) are machine-learning algorithms using similar methods on unlabelled data sets, based on image-pixel differences applicable for the residual classification of gene expression profiles. The GAN is best defined as two-competing neural networks wherein a generative model captures the data distribution while the discriminative model predicts the probability a sample data came from the training dataset rather than the generated dataset. Pertinent examples include CycleGAN and GibbsNet, iterative adversarial networks used for image-to-image translation inferences in complex datasets such as traffic flows (Zhu et al., 2017 [234]; Lamb et al., 2017 [235]). Saliency maps and other generative models are also used by Deep Learning networks to accomplish similar tasks in image processing and pattern recognition.

Moreover, the optimization algorithms (e.g., Bellman-Ford, Dijkstra's, fast marching) discussed in the graph-theoretic flow networks can be replaced by Hamilton-Jacobi-Bellman (HJB) equations pertaining to fluid models (Fürsikov et al., 2000 [236]). The solution of the Hamilton-Jacobi-Bellman equation is a partial differential equation that gives the optimal (minimum) cost flow for a *dynamical decision* problem. HJB is a necessary and sufficient condition to find optimal time paths (i.e., finding local minima/attractors) of control variables in feedback loop systems. It is analogous to the Hamilton-Jacobi equation with the energy term minimized with respect to a weight or control parameter $u(t)$, given as:

$$\frac{\partial V(q,t)}{\partial t} + \min_u H(q, p, t; u) = 0$$

where, $p = \nabla V$, $H$ is the analogous Hamiltonian (energy), and $q(t)$ is the state vector of the system. This is a complexity equation since min H adaptations are often (currently) intractable. Optimal control problems are generally nonlinear and without analytic solution. Therefore, HJB is often handled with stochastic methods as seen with the cost flow optimization algorithms discussed above assessing cell fate decision-making. Machine learning algorithms often treat fluid dynamics problems as regression or flow optimization control tasks (Brunton et al., 2019 [237]).

In principle, even simple binary classifiers like the multi-layered perceptron (MLP) provided by PyTorch, an open source Deep Learning framework, and machine learning platforms like Scikit-learn can fit the differential equation of a strange attractor as weights into an experimental dataset and study whether strange attractor patterns are observed. PyTorch neural networks use backpropagation and gradient descent learning to learn the parameters of the differential equations pertaining to the a priori assumed attractor model that fits the data.

Recurrent neural networks are sub-types of artificial neural networks that can accurately map the phase-space portraits of chaotic systems (Cestnik and Abel, 2019 [238]). While time-delay coordinate embedding is effective for the attractor reconstruction of low-dimensional systems, Reservoir Computing (RC) is a machine-learning algorithm that trains recurrent neural networks to find the Lyapunov exponents of high-dimensional datasets. For example, Reservoir Computing has demonstrated applicability in the chaotic attractor reconstruction of high-dimensional systems exhibiting spatio-temporal chaos such as complex fluid flows (Nakai and Saiki, 2018 [239]). The RC well predicted the short-term time-series forecasting of the Kuramoto-Sivashinsky (KS) equation to several multiples of Lyapunov time. The KS equation is a chaotic system whose pattern formation closely resembles that of fluid turbulence (Pathak et al., 2018 [240]; 2018 [241]). Hence, this is another class of machine learning algorithm that can be utilized on time-series cancer datasets to test the presence of strange attractors.

Furthermore, complex networks can be applied in the analysis of fluid dynamic structures and patterns (Gustafson and Hartman, 1985 [242]; Scarsoglio et al., 2016 [243]). Unsteady fluid flows can be visualized as graph-theoretic networks (Newman, 2003 [244]; Nair and Taira, 2015 [245]). Hopfield networks and ensemble Deep learning architectures can predict irregular patterns observed in complex fluid flows as well (Kutz et al., 2017 [246]; Yang and Huang, 2016 [247]). Strange attractors have been mapped in recurrent neural networks and gene networks (Mestl et al., 1996 [55]). Ling et al. (2015 [248]; 2016 [249]) first-demonstrated the applicability of Deep Learning Networks (DNN) to predict the turbulent flows of the Reynolds-averaged Navier-Stokes equations. A Galilean-invariance embedded, DNN network architecture (Tensor Basis Neural Network) underwent training on various turbulent flow datasets followed by the Bayesian optimization for the neural network's hyper-parameters (i.e., the number of hidden layers, the number of nodes per hidden layer, and the gradient descent algorithm's learning rate). Therefore, Deep Learning architectures trained for mapping complex fluid flows can be optimized for detecting strange attractors within cancer networks.

On a final note, the applicability of quantum computation in complexity theory remains at infancy. For example, quantum annealing is a method for finding solutions (global minima) to combinatorial optimisation problems and ground states of glassy systems. Quantum machine learning approaches have been used to classify and rank binding affinities of transcription factors, in regulating gene expression (Li et al., 2018 [250]). Quantum annealing transforms the Ising model Hamiltonian of a system into the Quadratic Unconstrained Binary Optimization (QUBO) problem, which is then solved by the DW2X processor (Li et al., 2018 [250]). Quantum approaches are very controversial and whether quantum mechanics has anything to contribute can only be verified in time.

**CONCLUSION**

To conclude, Complex systems theory provides a computational framework to cancer networks reconstruction and cancer reprogramming. Reconstructing cancer networks from a tumor sample's gene expression data and finding attractors in their state-space is an intractable (NP-complete) problem. The availability of efficient algorithms in detecting strange attractors in gene expression datasets remains scarce and troublesome. Artificial Intelligence (AI) algorithms are breakthrough approaches in cancer research. Graph-theoretic tools, bioinformatics and machine learning algorithms are currently exploited in the network science of cancer datasets. Data mining classification algorithms can distinguish cancer and healthy exosomes, paving early detection efforts. Deep Learning Networks and Reservoir Computing have demonstrated applications in chaotic time-series prediction, pattern recognition and classification of large complex datasets.

Algorithmic Information Dynamics (AID) is emerging as a more robust approach to the inference of causal relationships in cancer networks. AID is a powerful algorithmic framework which combines tools from Algorithmic Information Theory and perturbation analysis to study dynamical systems in software space. An application of which is the reconstruction of attractor landscapes from cancer datasets, such as those pertaining to the GBM stemness networks identified by Suvà et al. (2014). While currently employed machine learning algorithms in biological network reconstruction and pseudo-time ordering methods for gene expression datasets are based on statistical learning approaches, they do not inform us about *cause and effect* in the datasets. The causal structure inference of cell fate transitions during cancer stem cell differentiation (pattern formation) and their underlying GRN information dynamics are not well-captured by machine learning approaches currently used in cell lineage and network visualization. They only perform pattern recognition on statistical features of the datasets heavily dependent on dimensionality reduction techniques, and often fail to treat the cells and networks as *dynamical systems* (Zenil et al., 2019 [251]). As such, AID is a vastly unexplored artificial intelligence (AI) platform available for inferring causality in gene expression dynamics, cancer stem cell differentiation mapping and potentially map strange attractors in gene expression networks (Zenil et al., 2019 [251]). Furthermore, the implications of combining AID with ensemble Deep Learning architectures and RC computing remains to be explored given they are promising candidates for the detection of chaotic gene expression dynamics and strange attractor behaviors in Gene Regulatory Networks.

Numerous molecular techniques can resolve the technical challenges in mapping strange attractors within cancer datasets. For example, double cell-state reporters can map cell fate switches observed in cancer stem cells during reprogramming or exosome-mediated transitions in time-lapse video-microscopy and gene expression profiling experiments. Various unexplored machine learning algorithms and computational fluid models are further proposed as alternatives to detect strange attractors in cancer networks. However, generating time-series scRNA-Seq data remains both a technical and financial roadblock in computational oncology. A better understanding of the fluid dynamics governing cell fate transitions is proposed for more accurate reconstructions of cancer interactomes and forecasts of their gene expression dynamics.

**ACKNOWLEDGEMENTS:** Thanks to Dr. Phil Gold, Dr. Rolando Del Maestro, Dr. Goffredo Arena and Dr. Lorenzo Ferri of McGill University, and to Dr. Mario D'Amico of Concordia University for being my initial conditions.

**REFERENCES**

[1] D. Hanahan and R. A. Weinberg, "Hallmarks of Cancer: The Next Generation," *Cell*, **144**(5), 2011 pp. 646–674. https://doi.org/10.1016/j.cell.2011.02.013.

[2] J. Ladyman and K. Wiesner, *What Is a Complex System?*, New Haven: Yale University Press, 2020.

[3] C. R. Shalizi, "Methods and Techniques of Complex Systems Science: An Overview," *Complex Systems Science in Biomedicine* (T. S. Deisboeck and J. Y. Kresh, eds.), New York: Springer, 2006, pp. 33–114. https://doi.org/10.1007/978-0-387-33532-2_2.

[4] S. Wolfram, "Complex Systems Theory," in *Emerging Syntheses in Science: Proceedings of the Founding Workshops of the Santa Fe Institute, Santa Fe, New Mexico* (D. Pines, ed.), Redwood City, CA: Addison-Wesley, 1988 pp. 183–189.

[5] S. Wolfram, *A New Kind of Science*, Champaign, IL: Wolfram Media, Inc., 2002.

[6] C. Gros, *Complex and Adaptive Dynamical Systems: A Primer*, 2nd ed., Heidelberg: Springer, 2011.

[7] S. Wolfram, "*Cellular Automata as Models of Complexity*," *Nature* **311**, 1984 pp. 419-424. https://doi.org/10.1038/311419a0.

[8] Y. Jiao and S. Torquato, "*Emergent Behaviors from a Cellular Automaton Model for Invasive Tumor Growth in Heterogeneous Microenvironments*," PLoS Computational Biology **7**(12), 2011: e1002314. https://doi.org/10.1371/journal.pcbi.1002314.

[9] A. Monteagudo and J. Santos, *"A Cellular Automaton Model for Tumor Growth Simulation,"* In: *6th International Conference on Practical Applications of Computational Biology & Bioinformatics. Advances in Intelligent and Soft Computing* (M. Rocha, N. Luscombe, F. Fdez-Riverola, J. Rodríguez, eds.), Berlin, Heidelberg: Springer, **154**, 2012 pp. 147-155. https://doi.org/10.1007/978-3-642-28839-5_17

[10] D.S. Johnson and M. Garey, *Computers and Intractability: A Guide to the Theory of NP-Completeness,* New York: W.H. Freeman & Co., 1979.

[11] M. Sipser, *Introduction to the Theory of Computation*, Boston: PWS Publishing Co., 1997.

[12] P. Hou , Y. Li, X. Zhang, C. Liu, J. Guan, H. Li, T. Zhao et al. *"Pluripotent Stem Cells Induced from Mouse Somatic Cells by Small Molecule Compounds,"* Science **341**(6146), 2013 pp. 651-654. https://doi.org/10.1126/science.1239278.

[13] Y. Rais , A. Zviran, S. Geula, O. Gafni, E. Chomsky, S. Viukov, A. A. Mansour et al. *"Deterministic Direct Reprogramming of Somatic Cells to Pluripotency," Nature* **502** (7469), 2013 pp. 65-70. https://doi.org/10.1038/nature12587.

[14] S. Ronquist, G. Patterson, L. A. Muir, S. Lindsly, H. Chen, M. Brown, M. S. Wicha, A. Bloch, R. Brockett, and I. Rajapakse, "Algorithm for Cellular Reprogramming," *Proceedings of the National Academy of Sciences of the United States of America* **114**(45), 2017 pp. 11832-37. https://doi.org/10.1073/pnas.1712350114.

[15] C. Hernandez, Z. Wang, B. Ramazanov, Y. Tang, S. Mehta, C. Dambrot, Y. W. Lee, K. Tessema, I. Kumar, M. Astudillo, T. A. Neubert, S. Guo, N. B. & Ivanova, *"Dppa2/4 Facilitate Epigenetic Remodeling during Reprogramming to Pluripotency," Cell stem cell* **23**(3), 2018 pp. 396–411.e8. https://doi.org/10.1016/j.stem.2018.08.001.

[16] I. Kogut, S.M. McCarthy, M. Pavlova, D. P. Astling, X. Chen, A. Jakimenko, K.L. Jones, A. Getahun, J.C. Cambier, A.M.G. Pasmooij, M.F. Jonkman, D.R. Roop, and G. Bilousova, *"High-efficiency RNA-based Reprogramming of Human Primary Fibroblasts," Nature Communications* **9** (1):745, 2018. https://doi.org/10.1038/s41467-018-03190-3.

[17] S. Saito, Y.C. Lin, Y. Nakamura, R. Eckner, K. Wuputra, K.K. Kuo, C.S. Lin & K.K. Yokoyama, *"Potential Application of Cell Reprogramming techniques for Cancer Research," Cellular and Molecular Life Sciences: CMLS* **76**(1), 2019 pp. 45–65. https://doi.org/10.1007/s00018-018-2924-7.

[18] C. O'Brien-Ball & A. Biddle, "Reprogramming to Developmental Plasticity in Cancer Cells," Developmental Biology **430**(2), 2017 pp. 266-274. https://doi.org/10.1016/j.ydbio.2017.07.025.

[19] R. Khoshchehreh, M. Totonchi, J. Carlos Ramirez, R. Torres, H. Baharvand, A. Aicher, M. Ebrahimi & C. Heeschen, *"Epigenetic reprogramming of primary pancreatic cancer cells counteracts their in vivo tumourigenicity," Oncogene* **38**(34), 2019 pp. 6226–6239. https://doi.org/10.1038/s41388-019-0871-x.

[20] A. Carpentieri, E. Cozzoli, M. Scimeca, E. Bonanno, A.M. Sardanelli, & A. Gambacurta, *"Differentiation of human neuroblastoma cells toward the osteogenic lineage by mTOR inhibitor," Cell Death & Disease* **7**(4): e2202, 2016. https://doi.org/10.1038/cddis.2016.60.

[21] D. Bonnetand J.E. Dick, *"Human Acute Myeloid Leukemia is Organized as a Hierarchy that Originates from a Primitive Hematopoietic Cell," Nature Medicine* **3**(7), 1997 pp. 730–737. https://doi.org/10.1038/nm0797-730.

[22] L.I. Shlush, S. Zandi, A. Mitchell, W.C. Chen, J.M. Brandwein, V. Gupta, J.A. Kennedy et al. *"Identification of Pre-Leukaemic Haematopoietic Stem Cells in Acute Leukaemia," Nature* **506** (7488), 2014 pp. 328–333. https://doi.org/10.1038/nature13038.

[23] S.W. Ng, A. Mitchell, J.A. Kennedy, W.C. Chen, J. McLeod, N. Ibrahimova, A. Arruda et al. *"A 17-gene stemness score for rapid determination of risk in acute leukaemia," Nature* **540**(7633), 2016 pp. 433–437. https://doi.org/10.1038/nature20598.

[24] E. Battle & H. Clevers, *"Cancer Stem Cells Revisited,"* Nature Medicine **23**(10), 2017 pp. 1124-1134. https://doi.org/10.1038/nm.4409.

[25] D. Capper, D. Jones, M. Sill, V. Hovestadt, D. Schrimpf, D. Sturm, C. Koelsche et al. *"DNA Methylation-based Classification of Central Nervous System Tumours,"* Nature **555**(7697), 2018 pp. 469-474.

[26] M. Filbin M and M. Monje, *"Developmental Origins and Emerging Therapeutic Opportunities for Childhood Cancer, Nature Medicine* **25**(3), 2019 pp. 367–376. https://doi.org/10.1038/s41591-019-0383-9.

[27] E.R. Lechman, B. Gentner, S.W. Ng, E.M. Schoof, P. van Galen, J.A. Kennedy, S. Nucera et al. "miR-126 Regulates Distinct Self-Renewal Outcomes in Normal and Malignant Hematopoietic Stem Cells," *Cancer Cell* **29**(2), 2016 pp. 214–228. https://doi.org/10.1016/j.ccell.2015.12.011

[28] H.S. Venkatesh, W. Morishita, A.C. Geraghty, D. Silverbush, S.M. Gillespie, M. Arzt, L.T. Tam et al. *"Electrical and Synaptic Integration of Glioma into Neural Circuits,"* Nature **573**(7775), 2019 pp. 539–545. https://doi.org/10.1038/s41586-019-1563-y.

[29] M.L. Suvà, N. Riggi & B.E. Bernstein, *"Epigenetic Reprogramming in Cancer,"* Science **339**(6127), 2013 pp. 1567-1570. https://doi.org/10.1126/science.1230184.

[30] M.L. Suvà, E. Rheinbay, S.M. Gillespie, A.P. Patel, H. Wakimoto, S.D. Rabkin, N. Riggi et al. *"Reconstructing and reprogramming the tumor propagating potential of glioblastoma stem-like cells,"* Cells **157**(3), 2014 pp. 580–594. https://doi.org/10.1016/j.cell.2014.02.030.

[31] J. Sunayama, A. Sato, K. Matsuda, K. Tachibana, K. Suzuki, Y. Narita, S. Shibui et al. "*Dual blocking of mTor and PI3K elicits a prodifferentiation effect on glioblastoma stem-like cells,"* Neuro-Oncology **12**(12), 2015 pp. 1205–1219. https://doi.org/10.1093/neuonc/noq103.

[32] R. Salloum, M.K. McConechy, L.G. Mikael, C. Fuller, R. Drissi, M. DeWire, H. Nikbakht et al. *"Characterizing Temporal Genomic Heterogeneity in Pediatric High-grade Gliomas,"* Acta neuropathologica communications, **5**(1): 78, 2017. https://doi.org/10.1186/s40478-017-0479-8

[33] H. Bai, A.S. Harmancı, E.Z. Erson-Omay, J. Li, S. Coşkun, M. Simon, B. Krischek et al. "Integrated Genomic Characterization of IDH1-mutant Glioma Malignant Progression," *Nature Genetics* **48**(1), 2016 pp. 59–66. https://doi.org/10.1038/ng.3457.

[34] V. Körber, J. Yang, P. Barah, Y. Wu, D. Stichel, Z. Gu, M. Fletcher et al. "*Evolutionary Trajectories of IDH-wildtype Glioblastomas Reveal a Common Path of early Tumorigenesis Instigated Years Ahead of Initial Diagnosis,"* Cancer Cell **35**(4), 2019 pp. 692-704.e12. https://doi.org/10.1016/j.ccell.2019.02.007.

[35] L.F. Stead and R.G.W. Verhaak, *"Doomed from the TERT? A two-stage model of tumorigenesis in IDH-wild-type glioblastoma,"* Cancer Cell **35**(4), 2019 pp. 542-544. https://doi.org/10.1016/j.ccell.2019.03.009.

[36] P.A. Northcott, I. Buchhalter, A.S. Morrissy, V. Hovestadt, J. Weischenfeldt, T. Ehrenberger, S. Gröbner et al. "The whole-genome landscape of medulloblastoma subtypes," *Nature* **547**(7663), 2017 pp. 311–317. https://doi.org/10.1038/nature22973.

[37] M.C. Vladoiu, I. El-Hamamy, L.K. Donovan, H. Farooq, B.L. Holgado, Y. Sundaravadanam, V. Ramaswamy et al. *"Childhood cerebellar tumours mirror conserved fetal transcriptional programs,"* Nature **572**(7767), 2019 pp. 67–73. https://doi.org/10.1038/s41586-019-1158-7.

[38] E. Wingrove, Z.Z. Liu, K.D. Patel, A. Arnal-Estapé, W.L. Cai, M.A. Melnick, K. Politi, et al. *"Transcriptomic Hallmarks of Tumor Plasticity and Stromal Interactions in Brain Metastasis*," Cell reports **27**(4), 2019 pp. 1277–1292.e7. https://doi.org/10.1016/j.celrep.2019.03.085.

[39] G. MacLeod, D.A. Bozek, N. Rajakulendran, V. Monteiro, M. Ahmadi, Z. Steinhart, M. Kushida et al. "Genome-Wide CRISPR-Cas9 Screens Expose Genetic Vulnerabilities and Mechanisms of Temozolomide Sensitivity in Glioblastoma Stem Cells," *Cell reports*, **27**(3), 2019 pp. 971–986.e9. https://doi.org/10.1016/j.celrep.2019.03.047.

[40] L. Gong, Q. Yan, Y. Zhang, X. Fang, B. Liu & X. Guan, "Cancer Cell Reprogramming: a Promising Therapy Converting Malignancy to Benignity," *Cancer Communications* **39**(1):48, 2019. https://doi.org/10.1186/s40880-019-0393-5.

[41] S. Xiong, Y. Feng & L. Cheng, "Cellular Reprogramming as a Therapeutic Target in Cancer," *Trends in Cell Biology* **29**(8), 2019 pp. 623–634. https://doi.org/10.1016/j.tcb.2019.05.001

[42] C.H. Waddington, *"The Canalization of Development and the Inheritance of Acquired Characters," Nature* **150**, 1942 pp. 563–565. https://doi.org/10.1038/150563a0.

[43] S. Bhattacharya, Q. Zhang, & M.E. Andersen, *"A deterministic map of Waddington's Epigenetic Landscape for Cell Fate Specification," BMC Systems Biology* **5**: 85, 2011. https://doi.org/10.1186/1752-0509-5-85.

[44] J. Wang, K. Zhang, L. Xu and E. Wang, *"Quantifying the Waddington Landscape and Biological Path for Development and Differentiation," Proceedings of the National Academy of Sciences of the United States of America* 108 (20), 2011 pp. 8257-8262. https://doi.org/10.1073/pnas.1017017108.

[45] T.J. Perkins, E. Foxall, L. Glass & R. Edwards, *"A Scaling Law for Random Walks on Networks,"* Nature Communications **5**:5121, 2014. https://doi.org/10.1038/ncomms6121.

[46] J. Davila-Vederrain, J.C. Martinez-Garcia & E.R. Alvarez-Buylla *"Modelling the Epigenetic Attractors Landscape: Toward a Post-Genomic Mechanistic Understanding of Development," Frontiers in genetics* **6**:160, 2015. https://doi.org/10.3389/fgene.2015.00160

[47] M. B. Elowitz, A. J. Levine, E. D. Siggia, and P. S. Swain, "Stochastic gene expression in a single cell," Science **297** (5584), 2002 pp. 1183-1186. https://doi.org/10.1126/science.1070919.

[48] A-L. Barabási and Z.N. Oltvai, "Network Biology: Understanding the Cell's Functional Organization," Nature Review Genetics **5**(2), 2014 pp. 101-113. https://doi.org/10.1038/nrg1272.

[49] X. Zhu, M. Gerstein & M. Snyder, *"Getting Connected: Analysis and Principles of Biological Networks," Genes & Development* **21**(9), 2007 pp.1010–1024. https://doi.org/10.1101/gad.1528707.

[50] P.S. Swain, *"Lecture Notes on Stochastic Models in Systems Biology"*, arXiv:1607.07806 [q-bio.QM], 2016.

[51] T. Chen, H.L. He & G.M. Church, *"Modeling Gene Expression with Differential Equations," Pacific Symposium on Biocomputing. Pacific Symposium on Biocomputing*, 1999 pp. 29–40. PMID: 10380183.

*[52]* M. Cross and H. Greenside, *Pattern Formation and Dynamics in nonequilibrium systems*, Cambridge, UK; New *York: Cambridge University Press, 2009.*

[53] *Z. Cao and R. Grima, "Linear Mapping Approximation of Gene Regulatory Networks with Stochastic Dynamics," Nature Communications* **9**: 3305, 2018. https://doi.org/10.1038/s41467-018-05822-0.

[54] R. Metzler, E. Barkai, and J. Klafter, *"Deriving fractional Fokker-Planck equations from a Generalised Master Equation," Europhysics Letters* **4** (4), 1999 pp. 431-436. https://doi.org/10.1209/epl/i1999-00279-7.

[55] T. Mestl, C. Lemay and L. Glass, *"Chaos in High-dimensional Neural and Gene Networks,"* Physica D **98**(1), 1996 pp. 33-52. https://doi.org/10.1016/0167-2789(96)00086-3.

[56] C. Li and J. Wang, *"Quantifying Cell Fate Decisions for Differentiation and Reprogramming of a Human Stem Cell Network: Landscape and Biological Paths,"* PLoS Computational Biology **9**(8): e1003165, 2013. https://doi.org/10.1371/journal.pcbi.1003165.

[57] J. Wang, *"Landscape and Flux Theory of Nonequilibrium Dynamical Systems with Applications to Biology,"* Advances in Physics **64**(1), 2015 pp. 1-137. https://doi.org/10.1080/00018732.2015.1037068

[58] A. Palmisano, and C. Priami, *"Stochastic Simulation Algorithm,"* In: *Encyclopedia of Systems Biology* (W. Dubitzky, O. Wolkenhauer, K.H. Cho, H. Yokota, eds.), New York, NY: Springer, 2013. https://doi.org/10.1007/978-1-4419-9863-7_768.

[59] U. Herbach, A. Bonnaffoux, T. Espinasse, and O. Gandrillon, *"Inferring Gene Regulatory Networks from Single-cell Data: A Mechanistic Approach,"* BMC Systems Biology **11**:105, 2017. https://doi.org/10.1186/s12918-017-0487-0.

[60] Y.T. Lin and N. E. Buchler, *"Efficient analysis of Stochastic Gene Dynamics in the Non-adiabatic Regime using Piecewise Deterministic Markov Processes,"* Journal of the Royal Society, Interface **15**(138), 2018: 20170804. https://doi.org/10.1098/rsif.2017.0804.

[61] A-L. Barabási and M. Posfai, *Network Science*, Cambridge, UK: Cambridge University Press, 2016.

[62] S. Kauffman, *"Homeostasis and Differentiation in Random Genetic Control Networks,"* Nature **224** (5215), 1969 pp. 177–178. https://doi.org/10.1038/224177a0.

[63] G. Stoll, E. Viara, E. Barillot, and L. Calzone, *"Continuous Time Boolean Modeling for Biological Signaling: Application of Gillespie Algorithm,"* BMC Systems Biology **6**: 116, 2012. https://doi.org/10.1186/1752-0509-6-116.

[64] J. D. Schwab, S. D. Kühlwein, N. Ikonomi, M. Kühl, H. A. Kestler, *"Concepts in Boolean Network Modeling: What do they all mean?"* Computational and Structural Biotechnology Journal **18**, 2020 pp. 571-582. https://doi.org/10.1016/j.csbj.2020.03.001.

[65] H. Kim and H. Sayama, *"How Criticality of Gene Regulatory Networks Affects the Resulting Morphogenesis under Genetic Perturbations,"* Artificial Life **24** (2), 2018 pp.85-105. https://doi.org/10.1162/artl_a_00262.

[66] L. Glass, and C. Hill, *"Ordered and Disordered Dynamics in Random Networks,"* Europhysics Letters **41**(6), 1998 pp. 599-604. https://doi.org/10.1209/epl/i1998-00199-0.

[67] B. Derrida, and Y. Pomeau, *"Random Networks of Automata: A Simple Annealed Approximation,"* Europhysics Letters **1**(2), 1986 pp. 45–49. https://doi.org/10.1209/0295-5075/1/2/001.

[68] B. Luque, and R.V. Solé, *"Lyapunov Exponents in Random Boolean Networks,"* Physica A **284**(1-4), 2012 pp. 33–45. https://doi.org/10.1016/S0378-4371(00)00184-9.

[69] J. Cao, X. Qi, and H. Zhao, *"Modeling Gene Regulation Networks Using Ordinary Differential Equations,"* Methods in Molecular Biology **802**, 2012 pp. 185-97. https://doi.org/10.1007/978-1-61779-400-1_12.

[70] I. Albert, J. Thakar, S. Li, R. Zhang, & R. Albert, *"Boolean network simulations for life scientists,"* Source Code for Biology and Medicine **3**:16, 2008. https://doi.org/10.1186/1751-0473-3-16

[71] N.E. Buchler, U. Gerland, and T. Hwa, *"On Schemes of Combinatorial Transcription Logic," Proceedings of the National Academy of Sciences of the United States of America* **100**(9), 2003 pp. 5136–5141. https://doi.org/10.1073/pnas.0930314100.

[72] N.E. Buchler, U. Gerland, and T. Hwa., "Nonlinear Protein Degradation and the Function of Genetic Circuits," *Proceedings of the National Academy of Sciences of the United States of America* **102** (27), 2005 pp. 9559-9564. https://doi.org/10.1073/pnas.0409553102.

[73] L. Glass, and J. Pasternack, *"Stable Oscillations in Mathematical Models of Biological Control Systems," Journal of Mathematical Biology* **6**: 207, 1978. https://doi.org/10.1007/BF02547797.

[74] M. Sedhigi, and A.M. Sengupta, *"Epigenetic Chromatin Silencing: Bistability and Front Propagation," Physical Biology* **4** (4), 2007 pp. 246–255.

[75] S. Huang, *"Genomics, Complexity and Drug Discovery: Insights from Boolean Network Models of Cellular Regulation," Pharmacogenomics* **2**(3), 2001 pp. 203–222. https://doi.org/10.1517/14622416.2.3.203.

[76] T. Akutsu, S. Kuhara, O. Maruyama, and S. Miyano, *"A System for Identifying Genetic Networks from Gene Expression Patterns Produced by Gene Disruptions and Overexpressions," Genome Informatics, Workshop on Genome Informatics* **9**, 1998 pp. 151-160. PMID: 11072331.

[77] M. Milano, and A. Roli, "Solving the Satisfiability Problem Through Boolean Networks," In: AI*IA 99: Advances in Artificial Intelligence, Lecture Notes in Computer Science **1792** (E. Lamma and P. Mello, eds.), Berlin, Heidelberg: Springer, 2000. https://doi.org/10.1007/3-540-46238-4_7.

[78] E. Barillot, L. Calzone, P. Hupe, J-P. Vert, and Z. Vinovyev, *Computational Systems Biology of Cancer*, Boca Raton: CRC Press, Taylor & Francis Group, 2013.

[79] S. Lu, G. Mandava, G. Yan and X. Lu, *"An Exact Algorithm for Finding Cancer Driver Somatic Genome Alterations: the Weighted Mutually Exclusive Maximum Set Cover Problem," Algorithms for Molecular Biology* **11**:11, 2016. https://doi.org/10.1186/s13015-016-0073-9.

[80] M.C. Hajkarim, E. Upfal, and F. Vandin, *"Differentially Mutated Subnetworks Discovery," Algorithms for Molecular Biology* **14**: 10, 2019. https://doi.org/10.1186/s13015-019-0146-7

[81] J.B. Pollack, "The induction of dynamical recognizers," Machine learning **7**, 1991 pp. 227–252. https://doi.org/10.1007/BF00114845

[82] T.S. Cubitt, J. Eisert, and M.M. Wolf, *"Extracting Dynamical Equations from Experimental Data is NP Hard," Physical Review Letters* **108**, 120503, 2012. https://link.aps.org/doi/10.1103/PhysRevLett.108.120503.

[83] E. Ahmed, *"Fractals and Chaos in Cancer Models," International Journal of Theoretical Physics* **32**, 1993 pp. 353–355. https://doi.org/10.1007/BF00673722.

[84] S. Huang, and S. Kauffman, *"How to Escape the Cancer Attractor: Rationale and Limitations of Multi-target Drugs," Seminars in Cancer Biology* **23**(4), 2013 pp. 270-278. https://doi.org/10.1016/j.semcancer.2013.06.003.

[85] J.X. Zhou, A.O. Pisco, H. Qian, and S. Huang, *"Nonequilibrium Population Dynamics of Phenotype Conversion of Cancer Cells," PLoS One* **9**(12): e110714, 2014. https://doi.org/10.1371/journal.pone.0110714.

[86] Q. Li, A. Wennborg, E. Aurell, E. Dekel, J-Z. Zhou, Y. Xu, S. Huang, and I. Ernberg, "Dynamics Inside the Cancer Cell Attractor Reveal Cell Heterogeneity, Limits of Stability, and Escape," *Proceedings of the National Academy of Sciences of the United States of America* **113** (10), 2016 pp. 2672-2677. https://doi.org/10.1073/pnas.1519210113

[87] D. Stokić, R. Hanel, S. Thurner, *"Inflation of the edge of chaos in a simple model of gene interaction networks,"* Physical review. E, Statistical, nonlinear, and soft matter physics **77**(6 Pt 1), 061917, 2008. https://doi.org/10.1103/PhysRevE.77.061917.

[88] R. Hanel, M. Pöchacker, and S. Thurner, *"Living on the Edge of Chaos: Minimally Nonlinear Models of Genetic Regulatory Dynamics,"* Philosophical Transactions of the Royal Society A **368**, 2010 pp. 5583-5596. https://doi.org/10.1098/rsta.2010.0267.

[89] M.L Heltberg, S. Krishna, and M.H. Jensen, *"On Chaotic Dynamics in Transcription Factors and the Associated Effects in Differential Gene Regulation,"* Nature Communications **10**: 71, 2019. https://doi.org/10.1038/s41467-018-07932-1.

[90] Z. Zhang, W. Ye, Y. Qian, Z. Zheng, X. Huang, and G. Hu, *"Chaotic Motifs in Gene Regulatory Networks,"* PLoS One **7**(7): e39355, 2012. https://doi.org/10.1371/journal.pone.0039355.

[91] D. Toker, F.T. Sommer, and M. D'Esposito, "A Simple Method for Detecting Chaos in Nature," Communications Biology **3**: 11, 2020. https://doi.org/10.1038/s42003-019-0715-9.

[92] R. Smith, P.S. Stumpf, S.J. Ridden, A. Sim, S. Filippi, H.A. Harrington, & B.D. MacArthur, *"Nanog Fluctuations in Embryonic Stem Cells Highlight the Problem of Measurement in Cell Biology,"* Biophysical Journal **112**(12), 2017 pp. 2641–2652. https://doi.org/10.1016/j.bpj.2017.05.005

[93] F. Takens, "*Detecting Strange Attractors in Turbulence*", In: *Dynamical Systems and Turbulence, Warwick 1980, Lecture Notes in Mathematics* **898** (D. Rand and L.S. Young, eds.), Berlin, Heidelberg: Springer, 1980. https://doi.org/10.1007/BFb0091924.

[94] E.M. Posadas, S.R. Criley, and D.S. Coffey, *"Chaotic Oscillations in Cultured Cells: Rat Prostate Cancer,"* Cancer Research **56**(16), 1996 pp. 3682-3688. http://cancerres.aacrjournals.org/content/56/16/3682.

[95] A.M. Turing, *"The Chemical Basis of Morphogenesis,"* Philosophical Transactions of the Royal Society B: Biological Sciences **237** (641), 1952 pp. 37–72. https://doi.org/10.1098/rstb.1952.0012

[96] R.A. Gatenby and A. Gawlinki, *"A Reaction-Diffusion Model of Cancer Invasion,"* Cancer Research **56** (24), 1996 pp. 5745-5753. http:// https://cancerres.aacrjournals.org/content/56/24/5745.

[97] I. Ramis-Conde, M.A.J. Chaplain, and A.R.A. Anderson, *"Mathematical Modelling of Cancer Cell Invasion of Tissue. Mathematical and Computer Modelling* **47**(5-6), 2008 pp. 533-545. https://doi.org/10.1016/j.mcm.2007.02.034

[98] L. Tang, A.L., van de Ven, D. Guo, V. Andasari, V. Cristini, K.C. Li, & X. Zhou, *"Computational Modeling of 3D Tumor Growth and Angiogenesis for Chemotherapy Evaluation,"* PLoS One **9**(1), e83962, 2014. https://doi.org/10.1371/journal.pone.0083962.

[99] T.E. Yankeelov, V. Quaranta, V., K.J. Evans, & E.C. Rericha, *"Toward a Science of Tumor Forecasting for Clinical Oncology,"* Cancer Research **75**(6), 2015 pp. 918–923. https://doi.org/10.1158/0008-5472.CAN-14-2233.

[100] T.T. Ivancevic, M.J. Bottema, and L.C. Jain, *"A Theoretical Model of Chaotic Attractor in Tumor Growth and Metastasis,"* arXiv:0807.4272 [q-Bio], 2008.

[101] M. Itik & S.P. Banks, *"Chaos in a Three-dimensional Cancer Model,"* International Journal of Bifurcation and Chaos **20**(1), 2010 pp. 71 -79. https://doi.org/10.1142/S0218127410025417.

[102] C. Letellier, F. Denis, L.A. Aguirre, "What Can be Learned from a Chaotic Cancer Model?" Journal of Theoretical Biology **322**, 2013 pp. 7-16. https://doi.org/10.1016/j.jtbi.2013.01.003.

[103] S. Khajanchi, M. Perc and D. Ghosh, "The Influence of Time Delay in a Chaotic Cancer Model," *Chaos* **28** (10): 103101, 2018. https://doi.org/10.1063/1.5052496.

[104] I. Prigogine and I. Stengers, *Order out of Chaos: Man's New Dialogue with Nature*, New York: Bantam Books, 1984.

[105] D. Ruelle, *Turbulence, Strange Attractors and Chaos*, Singapore: World Scientific Series on Nonlinear Science Series A **16**, 1995.

[106] E.N. Lorenz, "Deterministic Nonperiodic Flow," *Journal of the Atmospheric Sciences* **20** (2), 1963 pp.130-141. https://doi.org/10.1175/1520-0469(1963)020<0130:DNF>2.0.CO;2.

[107] D. Ruelle, "Hydrodynamic Turbulence as a Problem in Nonequilibrium Statistical Mechanics," *Proceedings of the National Academy of Sciences of the United States of America* **109** (50), 2012 pp. 20344-20346. https://doi.org/10.1073/pnas.1218747109.

[108] D. Ruelle, "*Nonequilibrium Statistical Mechanics of Turbulence,*" *Journal of Statistical Physics* **157**(2), 2014 pp. 205-208. https://doi.org/10.1007/s10955-014-1086-8.

[109] J. Halatek & E. Frey, "Rethinking Pattern Formation in Reaction–Diffusion Systems," *Nature Physics* 14, 2018 pp. 507-514 (2018) https://doi.org/10.1038/s41567-017-0040-5.

[110] J. Denk, S. Kretschmer, J. Halatek, C. Hartl, P. Schwille, and E. Frey, "*MinE Conformational Switching Confers Robustness on Self-organized Min Protein Patterns,*" *Proceedings of the National Academy of Sciences of the United States of America* **115** (18), 2018 pp. 4553-4558. https://doi.org/10.1073/pnas.1719801115.

[111] P. Glock, F. Brauns, J. Halatek, E. Frey, and P. Schwille, "Design of Biochemical Pattern Forming Systems from Minimal Motifs," *eLife* **8**: e48646, 2019. https://doi.org/10.7554/eLife.48646.

[112] Y. Dang, D.A.J. Grundel, and H. Youk, "Cellular Dialogues: Cell-Cell Communication through Diffusible Molecules Yields Dynamic Spatial Patterns," *Cell Systems* **10**(1), 2019 pp. 82–98. https://doi.org/10.1016/j.cels.2019.12.001.

[113] H. Yan, M. Romero-López, L.I. Benitez, K. Di, H.B. Frieboes, C.C.W. Hughes, D.A. Bota, and J.S. Lowengrub, *"3D Mathematical Modeling of Glioblastoma Suggests That Transdifferentiated Vascular Endothelial Cells Mediate Resistance to Current Standard-of-Care Therapy,"* *Cancer Research* **77**(15), 2017 pp 4171-4184. https://doi.org/10.1158/0008-5472.CAN-16-3094.

[114] N. Vauchelet and E. Zatorska, "Incompressible limit of the Navier—Stokes model with a growth term," *Nonlinear Analysis* **163**, 2017 pp. 34-59. https://doi.org/10.1016/j.na.2017.07.003.

[115] M. C. Marchetti, J. F. Joanny, S. Ramaswamy, T. B. Liverpool, J. Prost, M. Rao, and R. A. Simha, "Hydrodynamics of Soft Active Matter," *Reviews of Modern Physics* **85**(3):1143, 2013. https://link.aps.org/doi/10.1103/RevModPhys.85.1143.

[116] T.E. Bate, E.J. Jarvis, M.E. Varney, and K-T. Wu, "Collective Dynamics of Microtubule-based 3D Active Fluids from Single Microtubules," *Soft Matter* **15**(25), 2019 pp. 5006-5016. https://doi.org/10.1039/c9sm00123a.

[117] M. James, W.J.T. Bos, and M. Wilczek, "Turbulence and turbulent Pattern Formation in a Minimal Model for Active Fluids," *Physical Review Fluids* **3**(6), 2018 pp. 061101. https://link.aps.org/doi/10.1103/PhysRevFluids.3.061101.

[118] B. Martinez-Pratt, J. Ignés-Mullol, J. Casademunt, and F. Sagués, *"Selection mechanism at the onset of active turbulence,"* Nature Physics **15**, 2019 pp. 362-366. https://doi.org/10.1038/s41567-018-0411-6.

[119] I.V. Kalgin & S.F. Chekmarev, *"Turbulent Phenomena in Protein Folding,"* Physical review. E, Statistical, Nonlinear, and Soft Matter Physics **83**(1), 2011 pp. 011920.

[120] V.A. Andryushchenko, and S.F. Chekmarev, *"On hydrodynamic interpretation of folding of an α-helical protein,"* Thermophysics and Aeromechanics **23**(6), 2016 pp 941–944. https://doi.org/10.1134/S0869864316060184.

[121] S.F. Chekmarev, "Protein Folding Dynamics in the Space of Experimentally Measured Variables: Turbulence Phenomena," Journal of Applied Mechanics and Technical Physics **59**(5), 2018 pp 827–833. https://doi.org/10.1134/S0021894418050085.

[122] A. Kourtidis, S.P. Ngok, P. Pulimeno, R.W. Feathers, L.R. Carpio, T.R. Baker, J.M. Carr, Distinct E-Cadherin based Complexes Regulate Cell Behaviour through miRNA Processing of Src and p120-Catenin Activity," *Nature Cell Biology* **17**(9), 2015 pp. 1145–1157. https://doi.org/10.1038/ncb3227.

[123] W.T. Coffey, "Self-Organization, Complexity and Chaos: The New Biology for Medicine," Nature Medicine **4**(8), 1998 pp. 882-885. https://doi.org/10.1038/nm0898-882.

[124] J.W. Baish and R.K. Jain, *"Fractals and Cancer,"* Cancer Research **60**(14), 2000 pp. 3683-3688.

[125] F.E. Lennon, G.C. Cianci, N.A. Cipriani, T.A. Hensing, H.J. Zhang, C.T. Chen, S.D. Murgu, E.E. Vokes, M.W. Vannier & R. Salgia, *"Lung Cancer-a Fractal Viewpoint,"* Nature reviews Clinical Oncology **12**(11), 2015 pp. 664–675. https://doi.org/10.1038/nrclinonc.2015.108.

[126] A. Brú, D. Casero, S. de Franciscis, M.A. Herrero, *"Fractal analysis and tumor Growth,"* Mathematical and Computer Modelling **47**(5–6), 2008 pp. 546-559. https://doi.org/10.1016/j.mcm.2007.02.033.

[127] K. Metze, R. Adam and J. B. Florindo, "*The Fractal Dimension of Chromatin- A Potential Molecular Marker for Carcinogenesis, Tumor Progression and Prognosis*," Expert Review of Molecular Diagnostics **4**(19), 2019 pp. 299-312. https://doi.org/10.1080/14737159.2019.1597707.

[128] A.Islam, S.M. Reza & K.M. Iftekharuddin, "*Multifractal Texture Estimation for Detection and Segmentation of Brain Tumors*," IEEE Transactions on Bio-medical Engineering **60**(11), 2013 pp. 3204–3215. https://doi.org/10.1109/TBME.2013.2271383.

[129] J. Vasiljevic, J. Pribic, K. Kanjer, W. Jonakowski, J. Sopta, D. Nikolic-Vukosavljevic, and M. Radulovic, *"Multifractal Analysis of Tumour Microscopic Images in the Prediction of Breast Cancer Chemotherapy Response,"* Biomedical Microdevices **17**(5):93, 2015. https://doi.org/10.1007/s10544-015-9995-0.

[130] A. Esteva, A. Robicquet, B. Ramsundar, V. Kuleshov, M. DePristo, K. Chou, C. Cui, G. Corrado, S. Thrun & J. Dean, *"A Guide to Deep Learning in Healthcare,"* Nature Medicine **25**(1), 2019 pp. 24–29. https://doi.org/10.1038/s41591-018-0316-z.

[131] E.J. Topol, *"High Performance Medicine: The Convergence of Human and Artificial Intelligence,"* Nature Medicine **25** (1), 2019 pp. 44-56. https://doi.org/10.1038/s41591-018-0300-7

[132] Z. Zhang, P. Chen, M. McGough, F. Xing, C. Wang, M. Bui, Y. Xie et al. *"Pathologist-level Interpretable Whole-slide Cancer Diagnosis with Deep Learning,"* Nature Machine Intelligence **1**, 2019 pp. 236-245.

[133] Silverbush, D. et al. *"Simultaneous Interpretation of Multi-omics Data Improves the Identification of Cancer Driver Modules," Cell Systems* **8**(5), 2019 pp. 456-466.e5. https://doi.org/10.1016/j.cels.2019.04.005.

[134] C. H. Huang, P.M. Chang, C.W. Hsu, C.Y. Huang & K.L. Ng, "*Drug repositioning for non-Small Cell Lung Cancer by Using Machine Learning Algorithms and Topological Graph Theory," BMC Bioinformatics* **17** *Suppl 1*:2, 2016. https://doi.org/10.1186/s12859-015-0845-0.

[135] M. Ali, T. Aittokallio, "*Machine Learning and Feature Selection for Drug Response Prediction in Precision Oncology Applications*," Biophysical Reviews **11**(1), 2019 pp. 31-39. https://doi.org/10.1007/s12551-018-0446-z.

[136] Y. LeCun, Y. Bengio and G. Hinton, *"Deep Learning,"* Nature **521**, 2015 pp. 436-444. https://doi.org/10.1038/nature14539.

[137] H. Shan, A. Padole, F. Homayounieh, U. Kruger, R. D. Khera, C. Nitiwarangkul, M. K. Kalra & G. Wang. *"Competitive Performance of a Modularized Deep Neural Network Compared to Commercial Algorithms for Low-dose CT Image Reconstruction,"* Nature *Machine Intelligence* **1**, 2019 pp. 269-276. https://doi.org/10.1038/s42256-019-0057-9.

[138] H. Stower, *"AI for Breast-Cancer Screening,"* Nature Medicine **26**(2): 163, 2020. https://doi.org/10.1038/s41591-020-0776-9.

[139] T.C. Hollon, B. Pandian, A.R. Adapa, E. Urias, A.V. Save, S.S.S, Khalsa, D.G. Eichberg, *"Near real-time intraoperative brain tumor diagnosis using stimulated Raman histology and deep neural networks,"* Nature Medicine **26**(1), 2020 pp. 52-58. https://doi.org/10.1038/s41591-019-0715-9.

[140] C. Pan, O. Schoppe, A. Parra-Damas, R. Cai, M.I. Todorov, G. Gondi, B. von Neubeck, B., "*Deep Learning Reveals Cancer Metastasis and Therapeutic Antibody Targeting in the Entire Body," Cell* 179(7), 2019 pp. 1661-1676.e19. https://doi.org/10.1016/j.cell.2019.11.013.

[141] W. Tong and R.B. Altman, *"High Precision Protein Functional Site Detection Using 3D Convolutional Neural Networks," Bioinformatics* **35**(9), 2009 pp. 1503-1512. https://doi.org/10.1093/bioinformatics/bty813.

[142] L. Zhao, H. Zhao, and H. Yan, *"Gene Expression Profiling of 1200 Pancreatic Ductal Adenocarcinoma Reveals Novel Subtypes," BMC Cancer* **18**(1):603, 2018. https://doi.org/10.1186/s12885-018-4546-8.

[143] J.N. Kather, A.T. Pearson, N. Halama, D. Jäger, J. Krause, S.H. Loosen, and A. Marx, "Deep Learning Can Predict Microsatellite Instability Directly from Histology in Gastrointestinal Cancer," Nature Medicine **25**(7), 2019 pp.1054–1056. https://doi.org/10.1038/s41591-019-0462-y.

[144] Z. Castello and H.G. Martin, "*A Machine Learning Approach to Predict Metabolic Pathway Dynamics from Time-series Multiomics Data*," Npj Systems Biology and Applications **4**:19, 2018. https://doi.org/10.1038/s41540-018-0054-3.

[145] A. Dirkse, A. Golebiewska, T. Buder, P.V. Nazarov, A. Muller, S. Poovathingal, N. Brons et al. *"Stem Cell-associated Heterogeneity in Glioblastoma Results from Intrinsic Tumor Plasticity Shaped by the Microenvironment," Nature Communications* **10**(1), 1787, 2019. https://doi.org/10.1038/s41467-019-09853-z.

[146]    C. Neftel, J. Laffy, M.G. Filbin, T. Hara, M.E. Shore, G.J. Rahme, A.R. Richman et al. "*An Integrative Model of Cellular States, Plasticity, and Genetics for Glioblastoma,*" Cell *178*(4), 2019 pp. 835–849.e21. https://doi.org/10.1016/j.cell.2019.06.024.

[147]    A.P. Patel, I. Tirosh, J.J. Trombetta, A.K. Shalek, S.M. Gillespie, H. Wakimoto, H., D.P. Cahill et al. *"Single-cell RNA-seq Highlights Intratumoral Heterogeneity in Primary Glioblastoma,"* Science **344**(6190), 2014 pp. 1396–1401. https://doi.org/10.1126/science.1254257.

[148]    J. Yuan, H. M. Levitin, V. Frattini, E. C. Bush, D. M. Boyett, J. Samanamud, M. Ceccarelli et al., *"Single-cell transcriptome analysis of lineage diversity in high-grade glioma,"* Genome Medicine **10**:157, 2018. https://doi.org/10.1186/s13073-018-0567-9.

[149]    C.P. Couturier, S. Ayyadhury, P.U. Le, J. Nadaf, J. Monlong, G. Riva, R. Allache et al., *"Single-cell RNA-seq Reveals that Glioblastoma Recapitulates Normal Brain Development,"* Nature Communications **11**(1), 3406, 2020. https://doi.org/10.1038/s41467-020-17186-5.

[150]    B. Pang, J. Xu, J. Hu, F. Guo, L. Wan, M. Cheng & L. Pang, *"Single-cell RNA-seq Reveals the Invasive Trajectory and Molecular Cascades Underlying Glioblastoma Progression,"* Molecular Oncology **13**(12), 2019 pp. 2588–2603. https://doi.org/10.1002/1878-0261.12569.

[151]    Y. Guo, X. Ji, J. Liu, D. Fan, Q. Zhou, C. Chen, W. Wang et al., *"Effects of Exosomes on Pre-metastatic Niche Formation in Tumors. Molecular Cancer* **18**:39, 2019. https://doi.org/10.1186/s12943-019-0995-1.

[152]    T. B. Steinbichler, J. Dudás, S. Skvortsov, U. Ganswindt, H. Riechelmann & I-I. Skvortsova, *"Therapy resistance mediated by exosomes,"* Molecular Cancer **18**:58, 2019. https://doi.org/10.1186/s12943-019-0970-x.

[153]    I. Keklikoglou, C. Cianciaruso, E. Güç, M.L. Squadrito, L.M. Spring, S. Tazzyman, L. Lambein, et al., "*Chemotherapy elicits pro-metastatic extracellular vesicles in breast cancer models,*" Nature Cell Biology **21**(2), 2019 pp. 190-202. https://doi.org/10.1038/s41556-018-0256-3.

[154]    G. Camussi, M.C. Deregibus, S. Bruno, C. Grange, V. Fonsato & C. Tetta, *"Exosome/Microvesicle-Mediated Epigenetic Reprogramming of Cells,"* American Journal of Cancer Research *1*(1), 2011 pp. 98–110.

[155]    S. Zhou, M. Abdouh, V. Arena, M. Arena, and G.O. Arena,*"Reprogramming Malignant Cancer Cells toward a Benign Phenotype following Exposure to Human Embryonic Stem Cell Microenvironment,"* PloS One **12**,1 e0169899, 2017. https://doi.org/10.1371/journal.pone.0169899.

[156]    M. Jermyn, K. Mok, J. Mercier, J. Desroches, J. Pichette, K. Saint-Arnaud, L. Bernstein, M.C. Guiot, K. Petrecca, & F. Leblond, *"Intraoperative brain cancer detection with Raman spectroscopy in humans,"* Science translational medicine **7**(274), 274ra19, 2015. https://doi.org/10.1126/scitranslmed.aaa2384.

[157]    M. Brusatori, G. Auner, T. Noh, L. Scarpace, B. Broadbent & S.N. Kalkanis, *"Intraoperative Raman Spectroscopy,"* Neurosurgery clinics of North America **28**(4), 2017 pp. 633–652. https://doi.org/10.1016/j.nec.2017.05.014.

[158]    J. Zhao, H. Lui, D.I. Mclean, H Zeng*, "Automated Autofluorescence Background Subtraction Algorithm for Biomedical Raman Spectroscopy,"* Applied Spectroscopy **61**(11), 2007 pp. 1225-1232. https://doi.org/10.1366/000370207782597003.

[159]   S.Y. Shen, R. Singhania, G. Fehringer, A. Chakravarthy, M. Roehrl, D. Chadwick, P.C. Zuzarte et al., "Sensitive tumor detection and classification using plasma cell-free DNA methylomes,"Nature **563** (7732), 2018 pp. 579-583. https://doi.org/10.1038/s41586-018-0703-0.

[160]   A. Gualerzi, S. Kooijmans, S. Niada, S, Picciolini, A.T. Brini, G. Camussi & M. Bedoni, *"Raman spectroscopy as a quick tool to assess purity of extracellular vesicle preparations and predict their functionality," Journal of extracellular vesicles*, **8**(1), 1568780, 2019. https://doi.org/10.1080/20013078.2019.1568780.

[161]   H. Shin, H. Jeong, J. Park, S. Hong, and Y. Choi, *"Correlation between Cancerous Exosomes and Protein Markers Based on Surface-Enhanced Raman Spectroscopy (SERS) and Principal Component Analysis (PCA)," ACS Sensors* **3**(12), 2018 pp. 2637–2643. https://doi.org/10.1021/acssensors.8b01047.

[162]   T. Rojalin, B. Phong, H.J. Koster, and R.P. Carney, *"Nanoplasmonic Approaches for Sensitive Detection and Molecular Characterization of Extracellular Vesicles," Frontiers in Chemistry* **7** 279, 2019. https://doi.org/10.3389/fchem.2019.00279.

[163]   J. Park, M. Hwang, B. Choi, H. Jeong, J.H. Jung, H.K. Kim, S. Hong, J.H. Park & Y. Choi, *"Exosome classification by Pattern analysis of surface-enhanced Raman spectroscopy data for lung cancer," Analytical Chemistry* **89**(12), 2017 pp. 6695–6701. https://doi.org/10.1021/acs.analchem.7b00911.

[164]   N. Banaei, A. Foley, J.M. Houghton, Y. Sun & B. Kim, B, *"Multiplex detection of pancreatic cancer biomarkers using a SERS-based immunoassay," Nanotechnology* **28**(45), 2017 pp. 455101. https://doi.org/10.1088/1361-6528/aa8e8c.

[165]   A. Zlotogorski-Hurvitz, B.Z. Dekel, D. Malonek, R. Yahalom, and M. Vered, *"FTIR-based spectrum of salivary exosomes coupled with computational-aided discriminating analysis in the diagnosis of oral cancer," Journal of Cancer Research and Clinical Oncology* **145**(3), 2019 pp. 685–694. https://doi.org/10.1007/s00432-018-02827-6.

[166]   A. Hoshino, H.S. Kim, L. Bojmar, K.E. Gyan, M. Cioffi, J. Hernandez, C.P. Zambirinis et al., "Extracellular Vesicle and Particle Biomarkers Define Multiple Human Cancers," *Cell* **182**(4), 2020 pp. 1044–1061.e18. https://doi.org/10.1016/j.cell.2020.07.009.

[167]   G. Kar, A. Gursoy & O. Keskin, *"Human Cancer Protein-Protein Interaction Network: A Structural Perspective," PLoS Computational Biology* **5**(12), e1000601, 2009. https://doi.org/10.1371/journal.pcbi.1000601.

[168]   M. AlQuraishi, G. Koytiger, A. Jenney, G. MacBeath, P.K. Sorger, *"A Multiscale Statistical Mechanical Framework Integrates Biophysical and Genomic Data to Assemble Cancer Networks," Nature Genetics* **46**(12), 2014 pp. 1363-1371. https://doi.org/10.1038/ng.3138.

[169]   H.F. Fumiã, M.L. Martins, "*Boolean Network Model for Cancer Pathways: Predicting Carcinogenesis and Targeted Therapy Outcomes," PLoS One* **8**(7): e69008, 2013. https://doi.org/10.1371/journal.pone.0069008.

[170]   Y. Li, D. Liu, T. Li, and Y. Zhu, "*Bayesian Differential Analysis of Gene Regulatory Networks Exploiting Genetic Perturbations," BMC Bioinformatics* **21**:12, 2020. https://doi.org/10.1186/s12859-019-3314-3.

[171]   D-Q. Zhang, C. Zhou, S-Z. Chen, Y. Yang, B. Shi, *"Identification of hub genes and pathways associated with bladder cancer based on co-expression network analysis," Oncology Letters* 14(1), 2017 pp. 1115–1122.

[172] R. Liu, W. Zhang, Z. Liu, and H. Zhou, *"Associating Transcriptional Modules with Colon Cancer Survival Through Weighted Gene Co-expression Network Analysis,"* BMC Genomics **18**(1), 361, 2017.https://doi.org/10.1186/s12864-017-3761-z.

[173] F. Vandin, E. Upfal, B.J. Raphael, *"Algorithms for Detecting Significantly Mutated Pathways in Cancer,"* Journal of Computational Biology: A Journal of Computational Molecular Cell Biology **18(**3), 2011 pp. 507–522. https://doi.org/10.1089/cmb.2010.0265

[174] W. Zhang, J. Chien, J. Yong, and R. Kuang, *"Network-based Machine Learning and Graph Theory Algorithms for Precision Oncology,"* Npj Precision Oncology **1**: 25, 2017. https://doi.org/10.1038/s41698-017-0029-7.

[175] S. Mallik and Z. Zhao, *"Graph- And Rule-Based Learning Algorithms: A Comprehensive Review of Their Applications for Cancer Type Classification and Prognosis Using Genomic Data,"* Briefings in bioinformatics **21**(2), 2020 pp. 368–394. https://doi.org/10.1093/bib/bby120.

[176] J.S. Bang, N.Y. Choi, M. Lee, K. Ko, Y.S. Park & K. Ko, *"Reprogramming of Cancer Cells into Induced Pluripotent Stem Cells Questioned,"* International journal of stem cells **12**(3), 2019 pp. 430–439. https://doi.org/10.15283/ijsc19067.

[177] D. Merico, D. Gfeller, G.D. Bader*, "How to Visually Interpret Biological Data Using Networks,"* Nature Biotechnology **27**(10), 2009 pp. 921-4. https://doi.org/10.1038/nbt.1567.

[178] A. Mehreen, and T. Aittokallio, *"Machine Learning and Feature Selection for Drug Response Prediction in Precision Oncology Applications,"* Biophysical Reviews **11**(1), 2019 pp. 31-39. https://doi.org/10.1007/s12551-018-0446-z.

[179] R. Jansen, H. Yu, D. Greenbaum, Y. Kluger, N.J. Krogan, S. Chung, A. Emili, M. Snyder, J.F. Greenblatt & M. Gerstein, *"A Bayesian Networks Approach for Predicting Protein-Protein Interactions from Genomic Data,"* Science **302**(5644), 2003 pp. 449–453.

[180] Y. Luo, J. Peng, and J. Ma, *"When Causal Inference Meets Deep Learning,"* Nature Machine Intelligence **2**, 2020 pp. 426-427. https://doi.org/10.1038/s42256-020-0218-x.

[181] Z. Sun, L. Chen, H. Xin, Y. Jiang, Q. Huang, A.R. Cillo, T. Tabib, et al., *"A Bayesian Mixture Model for Clustering Droplet-based single-cell Transcriptomic Data from Population Studies,"* Nature Communications **10**(1), 2019 pp. 1649. https://doi.org/10.1038/s41467-019-09639-3.

[182] F. K. Ahmad, S. Deris and N.H. Othman, *"The Inference of Breast Cancer Metastasis through Gene Regulatory Networks,"* Journal of Biomedical Informatics **45**(2), 2012 pp. 350-362. https://doi.org/10.1016/j.jbi.2011.11.015.

[183] K.W. Pajtler, H. Witt, M. Sill, D.T. Jones, V. Hovestadt, F. Kratochwil, K. Wani et al., *"Molecular Classification of Ependymal Tumors across All CNS Compartments, Histopathological Grades, and Age Groups,"* Cancer Cell **27**(5), 2015 pp. 728–743. https://doi.org/10.1016/j.ccell.2015.04.002.

[184] E. Gladilin, *"Graph-theoretical Model of Global Human Interactome Reveals Enhanced Long-range Communicability in Cancer Networks,"* PLoS One **12**(1): e0170953, 2017. https://doi.org/10.1371/journal.pone.0170953.

[185] P. Creixell, E.M. Schoof, J.T. Erler, and R. Linding, *"Navigating Cancer Network Attractors for Tumor-Specific Therapy,"* Nature Biotechnology **30**(9), 2012 pp. 842-848. https://doi.org/10.1038/nbt.2345

[186]     K. Basso, A.A. Margolin, G. Stolovitzky, U. Klein, R. Dalla-Favera & A. Califano, *"Reverse Engineering of Regulatory Networks in Human B Cells," Nature Genetics* **37**(4), 2005 pp. 382–390. https://doi.org/10.1038/ng1532.

[187]     S. Krishnaswamy, N. Zivanovic, R. Sharma, D. Pe'er & B. Bodenmiller, *"Learning Time-Varying Information Flow from Single-cell Epithelial to Mesenchymal Transition Data," PloS One* **13**(10), 2018 pp. e0203389. https://doi.org/10.1371/journal.pone.0203389.

[188]     S. Mohammadi, J. Davila-Velderrain, M. Kellis, *"Reconstruction of Cell-type-Specific Interactomes at Single-Cell Resolution," Cell Systems* **9**(6), 2019 pp. 559–568.e4. https://doi.org/10.1016/j.cels.2019.10.007.

[189]     A.E. Teschendorff and T. Enver, *"Single-cell Entropy for Accurate Estimation of Differentiation Potency from a Cell's Transcriptome," Nature Communications* **8**:15599, 2017. https://doi.org/10.1038/ncomms15599.

[190]     D.S. Fischer, A.K. Fiedler, E.M. Kernfeld, R. Genga, A. Bastidas-Ponce, M. Bakhti, H. Lickert, J. Hasenauer, R. Maehr & F.J. Theis, *"Inferring Population Dynamics from Single-cell RNA-Sequencing Time Series Data," Nature biotechnology* **37**(4), 2019 pp. 461–468. https://doi.org/10.1038/s41587-019-0088-0.

[191]     C. Lin and Z. Bar-Joseph, *"Continuous-state HMMs for Modeling Time-series Single-cell RNA-Seq Data," Bioinformatics* **35**(22), 2019 pp. 4707-4715. https://doi.org/10.1093/bioinformatics/btz296.

[192]     W. Saelens, R. Cannoodt, H. Todorov, Y. Saeys, *"A Comparison of Single-cell Trajectory Inference Methods. Nature Biotechnology* **37**(5), 2019 pp. 547-554. https://doi.org/10.1038/s41587-019-0071-9.

[193]     P. Yang, S.J. Humphrey, S. Cinghu, R. Pathania, A.J. Oldfield, D. Kumar, D. Perera, et al., "Multi-omic Profiling Reveals Dynamics of the Phased Progression of Pluripotency," *Cell Systems* **8**(5), 2019 pp. 427–445.e10. https://doi.org/10.1016/j.cels.2019.03.012.

[194]     T.E. Chan, M.P.H. Stumpf, and A.C. Babtie, "Gene Regulatory Network Inference from Single Cell Data using Multivariate Information Measures," Cell Systems **5**(3), 2017 pp. 251-267.e3. https://doi.org/10.1016/j.cels.2017.08.014

[195]     Y. Gan, N. Li, G. Zou, Y. Xin, J. Guan, *"Identification of Cancer Subtypes from Single-cell RNA-seq Data using a Consensus Clustering Method. BMC Medical Genomics* **11**: 117, 2018. https://doi.org/10.1186/s12920-018-0433-z.

[196]     K. Ozturk, M. Dow, D.E. Carlin, R. Bejar & H. Carter, *"The Emerging Potential for Network Analysis to Inform Precision Cancer Medicine," Journal of Molecular Biology*, **430**(18 Pt A), 2018 pp. 2875–2899. https://doi.org/10.1016/j.jmb.2018.06.016

[197]     M. Zwiessele and N.D. Lawrence, *"Topslam: Waddington Landscape Recovery for Single Cell Experiments,"* BioRxiv, 2017. https://doi.org/10.1101/057778

[198]     A. Butler, P. Hoffman, P. Smibert, E. Papalexi, R. Satija, *"Integrating single-cell transcriptomic data across different conditions, technologies and species," Nature Biotechnology* **36**(5), 2018 pp. 411-420. https://doi.org/10.1038/nbt.4096.

[199]     T. Stuart, A. Butler, P. Hoffman, C. Hafemeister, E. Papalexi, W.M. Mauck, Y. Hao, M. Stoeckius, P. Smibert & R. Satija, *"Comprehensive Integration of Single-Cell Data," Cell* **177**(7), 2019 pp. 1888–1902.e21. https://doi.org/10.1016/j.cell.2019.05.031.

[200]     F.A. Wolf, P. Angerer, F.J. Theis, *"SCANPY: Large-Scale Single Cell Gene Expression Data Analysis," Genome Biology* **19**(1):15, 2018. https://doi.org/10.1186/s13059-017-1382-0.

[201] K.R. Moon, D. van Dijk, Z. Wang, S. Gigante, D.B. Burkhardt, W.S. Chen, K. Yim et al., *"Visualizing Structure and Transitions in High-dimensional Biological Data,"* Nature Biotechnology **37**(12), 2019 pp. 1482-1492. https://doi.org/10.1038/s41587-019-0336-3.

[202] M. Pájaro, I. Otero-Muras, C. Vasquez, and A.A. Alonso, *"SELANSI: A Toolbox for Simulation of Stochastic Gene Regulatory Networks,"* Bioinformatics **34**(5), 2017 pp. 893–895. https://doi.org/10.1093/bioinformatics/btx645.

[203] S. Tripathi, J. Lloyd-Price, A. Ribeiro, O. Yli-Harja, M. Dehmer, F. Emmert-Streib, *"sgnesR: An R package for Simulating Gene Expression Data from an Underlying Real Gene Network Structure Considering Delay Parameters,"* BMC Bioinformatics **18**: 325, 2017. https://doi.org/10.1186/s12859-017-1731-8.

[204] E. Lummertz da Rocha, R.G. Rowe, V. Lundin, M. Malleshaiah, D.K. Jha, C.R. Rambo, H. Li, T.E. North, J.J. Collins G.Q. & Daley, *"Reconstruction of complex single-cell trajectories using CellRouter,"* Nature Communications **9**(1), 892, 2018. https://doi.org/10.1038/s41467-018-03214-y.

[205] S. Jin, A.L. MacLean, T. Peng, Q. Nie, *"scEpath: Energy Landscape-based Inference of Transition Probabilities and Cellular Trajectories from Single-cell Transcriptomic Data,"* Bioinformatics **34**(12), 2018 pp. 2077-2086. https://doi.org/10.1093/bioinformatics/bty058.

[206] K. Street, D. Risso, R.B. Fletcher, D. Das, J. Ngai, N. Yosef, E. Purdom & S. Dudoit, *"Slingshot: Cell Lineage and Pseudotime Inference for Single-cell Transcriptomics,"* BMC Genomics **19**(1), 477, 2018. https://doi.org/10.1186/s12864-018-4772-0.

[207] J. Shi, A.E. Teschendorff, W. Chen, L. Chen, T. Li, *"Quantifying Waddington's Epigenetic Landscape: A Comparison of Single-Cell Potency Measures,"* Briefings in Bioinformatics **21**(1), 2020 pp. 248–261. https://doi.org/10.1093/bib/bby093.

[208] K.L. Buschur, M. Chikina, P. V Benos, *"Causal network perturbations for instance-specific analysis of single cell and disease samples,"* Bioinformatics **36**(8), 2020 pp. 2515–2521. https://doi.org/10.1093/bioinformatics/btz949.

[209] J.J. Hopfield, *"Neural Networks and Physical Systems with Emergent Collective Computational Abilities,"* Proceedings of the National Academy of Sciences of the United States of America **79**(8), 1982 pp. 2554–2558. https://doi.org/10.1073/pnas.79.8.2554.

[210] J.J. Hopfield, D.W. Tank, "Neural Computation of Decisions in Optimization Problems," Biological Cybernetics *52*(3), 1985 pp. 141–152. https://doi.org/10.1007/BF00339943.

[211] T.R.J. Bossomaier and D.J. Green, *Complex Systems*, Cambridge, UK: Cambridge University Press, 2000.

[212] A. Szedlak, G. Paternostro, C. Piermarocchi, *"Control of Asymmetric Hopfield Networks and Application to Cancer Attractors,"* PLoS One **9**(8): e105842, 2014. https://doi.org/10.1371/journal.pone.0105842.

[213] A. H. Lang, H. Li, J.J. Collins, and P. Mehta, *"Epigenetic Landscapes Explain Partially Reprogrammed Cells and Identify Key Reprogramming Genes,"* PLOS Computational Biology **10**(8): e1003734, 2014. https://doi.org/10.1371/journal.pcbi.1003734.

[214] J. Guo, and J. Zheng, *"Hopland: Single Cell Pseudotime Recovery using Continuous Hopfield Network-based Modelling of Waddington's Epigenetic Landscape,"* Bioinformatics **33**(14), 2017 pp. i102-109. https://doi.org/10.1093/bioinformatics/btx232.

[215] A.J. Conforte, L. Alves, F.C. Coelho, N. Carels & F. da Silva, *"Modeling Basins of Attraction for Breast Cancer Using Hopfield Networks,"* Frontiers in Genetics **11**: 314, 2020. https://doi.org/10.3389/fgene.2020.00314.

[216] J.K. Grewal, B. Tessier-Cloutier, M. Jones, S. Gakkhar, Y. Ma, R. Moore, A.J. Mungall et al., *"Application of a Neural Network Whole Transcriptome-Based Pan-Cancer Method for Diagnosis of Primary and Metastatic Cancers,"* JAMA Network Open **2**(4), e192597, 2019. https://doi.org/10.1001/jamanetworkopen.2019.2597

[217] K. Kourou, T.P. Exarchos, K.P. Exarchos, M.V. Karamouzis & D.I. Fotiadis, "Machine Learning Applications in Cancer Prognosis and Prediction," *Computational and Structural Biotechnology Journal* **13**, 2015 pp. 8–17. https://doi.org/10.1016/j.csbj.2014.11.005.

[218] T. Tian, J. Wan, Q. Song, Z. Wei, *"Clustering Single-cell RNA-seq Data with a Model-based Deep Learning Approach,"* Nature Machine Intelligence **1**, 2019 pp. 191–198. https://doi.org/10.1038/s42256-019-0037-0.

[219] S. Tasaki, C. Gaiteri, S. Mostafavi, Y. Wang, *"Deep learning Decodes the Principles of Differential Gene Expression,"* Nature Machine Intelligence **2**, 2020 pp. 376-386. https://doi.org/10.1038/s42256-020-0201-6

[220] M. Amodio, D. van Dijk, K. Srinivasan, W.S. Chen, H. Mohsen, K.R. Moon, A. Campbell et al., *"Exploring single-cell data with Deep Multitasking Neural Networks,"* Nature Methods **16**, 2019 pp. 1139-1145. https://doi.org/10.1038/s41592-019-0576-7.

[221] D. Aguilar-Hidalgo, M.C. Lemos, and A. Cordoba, *"Evolutionary Dynamics in Gene Networks and Inference Algorithms,"* Computation **3**(1), 2015 pp. 99-113. https://doi.org/10.3390/computation3010099.

[222] M.S. Krieger, J.M. Moreau, H. Zhang, M. Chien, J.L. Zehnder, M.A. Nowak, and M. Craig, *"Novel Cytokine Interactions Identified During Perturbed Hematopoiesis,"* BioRxiv, 2018. https://doi.org/10.1101/484170.

[223] H. Zenil, N.A. Kiani, F. Marabita, Y. Deng, S. Elias, A. Schmidt, G. Ball and J. Tegnér, *"An Algorithmic Information Calculus for Causal Discovery and Reprogramming Systems,"* iScience **19**, 2019 pp. 1160-1172. https://doi.org/10.1016/j.isci.2019.07.043.

[224] H. Zenil, N.A. Kiani, F.S. Abrahao, J. Tegnér, *"Algorithmic Information Dynamics,"* *Scholarpedia* 15(7):53143, 2020. https:// doi:10.4249/scholarpedia.53143.

[225] H. Zenil, A. Schmidt, and J. Tegnér, *"Causality, Information and Biological Computation: An Algorithmic Software Approach to Life, Disease and the Immune System,"* In: *From matter to life: information to causality*, Part III: Chapter 11, pp. 244–280. (S. I. Walker, P. C.W. Davies and G. Ellis, eds.), Cambridge University Press, 2017. https://doi.org/10.1017/9781316584200.

[226] O.E. Lanford III, *"The Strange Attractor Theory of Turbulence,"* Annual Reviews of Fluid Mechanics **14**, 1982 pp. 347-364. https://doi.org/10.1146/annurev.fl.14.010182.002023.

[227] F. Soler-Toscano, H. Zenil, J-P. Delahaye, N. Gauvrit, *"Calculating Kolmogorov Complexity from the Output Frequency Distributions of Small Turing Machines,"* PLoS One **9**(5): e96223, 2014. https://doi.org/10.1371/journal.pone.0096223.

[228] H. Zenil, N.A. Kiani and J. Tegnér, *"Methods of Information Theory and Algorithmic Complexity for Network Biology,"* Seminars in Cell & Developmental Biology **51**, 2016 pp. 32-43. https://doi.org/10.1016/j.semcdb.2016.01.011.

[229] H. Zenil, N.A. Kiani and J. Tegnér, "*A Review of Graph and Network Complexity from an Algorithmic Information Perspective,*" *Entropy* **20**(8): 551, 2018. https://doi.org/10.3390/e20080551.

[230] I. Ayed, E. de Bezenac, A. Pajot & P. Gallinari, *"Learning the Spatio-Temporal Dynamics of Physical Processes from Partial Observations,"* ICASSP 2020 - 2020 IEEE International Conference on Acoustics, Speech and Signal Processing (ICASSP). DOI: 10.1109/ICASSP40776.2020.9053035

[231] V.S. Lempitsky and A. Zisserman, *"Learning to count objects in images,"* In: NIPS'10 *Proceedings of the 23rd International Conference on Neural Information Processing Systems* **1**, 2010 pp. 1324-1332. http://papers.nips.cc/paper/4043-learning-to-count-objects-in-images.

[232] J.P. Cohen, G. Boucher, C.A. Glastonbury, H.Z. Lo, Y. Bengio, "*Count-Ception: Counting by Fully Convolutional Redundant Counting,*" *2017 IEEE International Conference on Computer Vision Workshops (ICCVW)*, 2018 pp. 18–26. https://doi.org/10.1109/ICCVW.2017.9.

[233] E. Lu, W. Xie, and A. Zisserman, *"Class-Agnostic Counting,"* Asian Conference on Computer Vision (ACCV), 2018.

[234] J-Y. Zhu, T. Park, P. Isola, A.A. Efros, "*Unpaired Image-to-Image Translation using Cycle-Consistent Adversarial Networks*", In: *IEEE International Conference on Computer Vision (ICCV),* 2017. http://doi.ieeecomputersociety.org/10.1109/ICCV.2017.244.

[235] A. Lamb, D. Hjelm, Y. Ganin, J.P. Cohen, A.C. Courville, Y. Bengio. *"GibbsNet: Iterative Adversarial Inference for Deep Graphical Models,"* In: *Advances in Neural Information Processing Systems* **30** *(NIPS 2017)*, 2017.

[236] A. Fürsikov, M. Gunzburger, L.S. Hou, S. Manservisi, *"Optimal Control Problems for the Navier-Stokes Equations,"* In: *Lectures on Applied Mathematics* (H.J. Bungartz, R.H.W. Hoppe and C. Zenger, eds), Berlin, Heidelberg: Springer, 2000. pp 143-155. https://doi.org/10.1007/978-3-642-59709-1_11.

[237] S.L. Brunton, B.R. Noack, and P. Koumoutsakos, *"Machine Learning for Fluid Dynamics,"* Annual Reviews of Fluid Mechanics **52**, 2020 pp. 477-508. https://doi.org/10.1146/annurev-fluid-010719-060214.

[238] R. Cestnik and M. Abel, *"Inferring Dynamics of Oscillatory Systems using Recurrent Neural Networks,"* Chaos **29**, 063128, 2019. https://doi.org/10.1063/1.5096918

[239] K. Nakai and Y. Saiki, *"Machine-Learning Inference of Fluid Variables from Data using Reservoir Computing,"* Physical Review E **98**(2), 023111, 2018. https://link.aps.org/doi/10.1103/PhysRevE.98.023111.

[240] J. Pathkak, A. Wikner, R. Fussell, S. Chandra, B.R. Hunt, M. Girvan and E. Ott, *"Hybrid Forecasting of Chaotic Processes: Using Machine Learning in Conjunction with a Knowledge-based Model,"* Chaos **28**(4), 041101, 2018. https://doi.org/10.1063/1.5028373.

[241] J. Pathkak, B.R. Hunt, M. Girvan, Z. Lu and E. Ott, *"Model-free Prediction of Larger Spatiotemporally Chaotic Systems from Data: A Reservoir Computing Approach,"* Physical Review Letters **120**(2), 024102, 2018. https://doi.org/10.1103/PhysRevLett.120.024102.

[242] K. Gustafson and R, Hartman, *"Graph Theory and Fluid Dynamics,"* SIAM Journal on Algebraic and Discrete Methods **6**(4), 1985 pp. 643–656. https://doi.org/10.1137/0606064.

[243] S. Scarsoglio, G. Iacobello, and L. Ridolfi, *"Complex Networks Unveiling Spatial Patterns in Turbulence,"* International Journal of Bifurcation and Chaos **26** (13), 1650223, 2016. https://doi.org/10.1142/S0218127416502230.

[244]     M.E.J. Newman, *"The Structure and Function of Complex Networks,"* SIAM Review **45**(2), 2003 pp. 167-256. https://doi.org/10.1137/S003614450342480.

[245]     A.G. Nair, and K. Taira, "*Network-Theoretic Approach to Sparsified Discrete Vortex Dynamics,"* Journal of Fluid Mechanics **768**, 2015 pp. 549-571. https://doi.org/10.1017/jfm.2015.97.

[246]     J.N. Kutz, *"Deep Learning in Fluid Dynamics,"* Journal of Fluid Mechanics **814**, 2017 pp. 1-4. https://doi.org/10.1017/jfm.2016.803.

[247]     S. Yang and Y. Huang, *"Complex Dynamics in Simple Hopfield Neural Networks,"* Chaos 16, 033114, 2016. https://doi.org/10.1063/1.2220476.

[248]     J. Ling and J. Templeton, "*Evaluation of Machine Learning Algorithms for Prediction of Regions of High Reynolds Averaged Navier Stokes Uncertainty,"* Physics of Fluids **27**, 085103, 2015. https://doi.org/10.1063/1.4927765.

[249]     J. Ling*,* A. Kurzawski, J. Templeton, *"Reynolds Averaged Turbulence Modelling using Deep Neural Networks with Embedded Invariance,"* Journal of Fluid Mechanics **807**, 2016, pp. 155–166. https://doi.org/10.1017/jfm.2016.615.

[250]     R.Y. Li, R. Di Felice, R. Rohs, D.A. Lidar, *"Quantum Annealing versus Classical Machine Learning Applied to a Simplified Computational Biology Problem,"* Npj Quantum Information **4***:14*, 2018. https://doi.org/10.1038/s41534-018-0060-8.

[251]     H. Zenil, N.A. Kiani, A.A. Zea, and J. Tegnér, *"Causal Deconvolution by Algorithmic Generative Methods,"* Nature Machine Intelligence **1**, 2019 pp. 58–66. https://doi.org/10.1038/s42256-018-0005-0.